\documentclass[lscape]{emulateapj}
\usepackage{graphicx}
\usepackage{epstopdf}
\usepackage{color}
\usepackage{lscape}

\newcommand{\Mjup}{\mbox{$M_\mathrm{Jup}$}}
\newcommand{\Msun}{\mbox{$M_{\odot}$}}

\begin{document}
\shorttitle{Spectroscopic Confirmation of Young PMCs}
\title{Spectroscopic Confirmation of Young Planetary-Mass Companions \\ on Wide Orbits}
\author{Brendan P. Bowler,\altaffilmark{1,2,3,4} 
Michael C. Liu,\altaffilmark{2} 
Adam L. Kraus,\altaffilmark{5} 
Andrew W. Mann\altaffilmark{2,5,6} \\ }
\email{bpbowler@caltech.edu}

\altaffiltext{1}{California Institute of Technology, Division of Geological and Planetary Sciences, 1200 E. California Blvd., Pasadena, CA 91101 USA.}
\altaffiltext{2}{Institute for Astronomy, University of Hawai`i; 2680 Woodlawn Drive, Honolulu, HI 96822, USA.}
\altaffiltext{3}{Visiting Astronomer at the Infrared Telescope Facility, which is operated by the University of Hawaii under Cooperative Agreement no. NNX-08AE38A with the National Aeronautics and Space Administration, Science Mission Directorate, Planetary Astronomy Program.}
\altaffiltext{4}{Caltech Joint Center for Planetary Astronomy Fellow.}
\altaffiltext{5}{University of Texas at Austin, Astronomy Department, Austin, TX 78712, USA.}
\altaffiltext{6}{Harlan J. Smith Fellow.}
\altaffiltext{*}{Some of the data presented herein were obtained at the W.M. Keck Observatory, which is operated as a scientific partnership among the California Institute of Technology, the University of California and the National Aeronautics and Space Administration. The Observatory was made possible by the generous financial support of the W.M. Keck Foundation.}

\submitted{ApJ, accepted (Jan 28 2014)}

\begin{abstract}
We present  
moderate-resolution ($R$$\sim$4000--5000)  
near-infrared integral field spectroscopy 
of the young (1--5~Myr) 6--14~\Mjup \ companions ROXs~42B~b and FW~Tau~b 
obtained with  Keck/OSIRIS and Gemini-North/NIFS.
The spectrum of ROXs~42B~b exhibits clear signs of 
low surface gravity common to young L dwarfs,
confirming its extreme youth, cool temperature, and low mass.
Overall, it closely resembles 
the free-floating 
4--7~\Mjup \ L-type Taurus member 2MASS~J04373705+2331080.
The companion to FW~Tau~AB is more enigmatic.
Our optical and near-infrared spectra show 
strong evidence of outflow activity and disk accretion in the form of 
line emission from 
$[$\ion{S}{2}$]$, $[$\ion{O}{1}$]$, H$\alpha$, Ca~II, 
$[$\ion{Fe}{2}$]$, Pa$\beta$, and H$_2$.
The molecular hydrogen emission is spatially resolved 
as a single lobe that stretches $\approx$0$\farcs$1 (15~AU).
Although the extended emission is not kinematically resolved in our data, its morphology resembles 
shock-excited H$_2$ jets primarily seen in young Class~0 and Class~I sources.
The near-infrared continuum of FW~Tau~b is mostly flat and lacks
the deep absorption features expected for a cool, late-type object.
This may be a result of accretion-induced veiling, especially in light of its
strong and sustained H$\alpha$~emission ($EW$(H$\alpha$)$\gtrsim$290~\AA).  
Alternatively,  FW~Tau~b may  
be a slightly warmer (M5-M8) accreting low-mass star or brown dwarf (0.03--0.15~\Msun) with an edge-on disk.
Regardless, its young evolutionary stage is 
in stark contrast to its Class~III host FW~Tau~AB, indicating a more rapid disk
clearing timescale for the host binary system than for its wide companion.
Finally, we present near-infrared spectra of the young ($\sim$2--10~Myr) 
low-mass (12--15~\Mjup) companions GSC~6214-210~B and SR~12~C and find they 
best resemble low gravity M9.5 and M9 substellar templates.

\end{abstract}

\section{Introduction}{\label{sec:intro}}

Adaptive optics (AO) imaging searches for self-luminous giant planets are uncovering
a population of companions with separations of several hundred AU 
and masses near and below the deuterium-burning limit 
($\sim$13~\Mjup; e.g., \citealt{Spiegel:2011p22104}).
Together with  directly imaged planets
located at more moderate separations of $\sim$10--100~AU
(\citealt{Marois:2008p18841}; \citealt{Marois:2010p21591}; 
\citealt{Lagrange:2010p21645}; \citealt{Kraus:2012p23492}; \citealt{Kuzuhara:2013p25347}; \citealt{Rameau:2013ds}),
planetary-mass companions (PMCs) on wide
orbits are extending the scale of planetary systems to over five orders
of magnitude in separation ($\sim$0.1--1000~AU).
Altogether, a total of 12 PMCs to stars 
with projected separations in excess of 100~AU 
and masses $\lesssim$15~\Mjup \ have been found (Table~\ref{tab:pmc}).
With the exception of Ross~458~C and WD~0806-661~B, all of these systems are very young (most being $\lesssim$10~Myr)
and the companion spectral types span late-M to mid-L.  
This selection for extreme youth is primarily a result of
brown dwarf and giant planet evolution;
by 1 Gyr, 10 \Mjup \ objects cool to a mere 500~K 
(\citealt{Burrows:1997p2706}; \citealt{Baraffe:2003p587}),
placing them at the boundary between T and Y dwarfs (\citealt{Cushing:2011p23757})
where near-infrared (NIR) absolute magnitudes plummet
(\citealt{Kirkpatrick:2011p23642}; \citealt{Liu:2011p22852}; \citealt{Liu:2012p24074}; \citealt{Kirkpatrick:2012p24092}).  At 5~Myr, however, the same objects 
have effective temperatures of $\sim$2000~K and absolute 
magnitudes of $\sim$10--11~mag in the NIR
assuming  ``hot start'' formation (\citealt{Marley:2007p18269}), 
making them easy targets for 
NIR AO imaging surveys of 
star-forming regions (e.g., \citealt{Ireland:2011p21592}).

The existence of these wide-separation PMCs poses a challenge for models of giant planet formation.
Their orbital separations are far too large for core accretion to operate (e.g., 
\citealt{Pollack:1996p19730}; \citealt{Alibert:2005p17987}; \citealt{DodsonRobinson:2009p19734})
and models of disk instability 
are most efficient at smaller separations of
tens of AU where protoplanetary disks are both massive and cold  
enough to gravitationally collapse (\citealt{Boss:1997p18822}; \citealt{Mayer:2002p22604}; 
\citealt{Kratter:2010p20098}; \citealt{Boss:2011p22047}).
Other formation scenarios include the direct fragmentation of molecular cloud cores
(e.g., \citealt{Bate:2002p20246}; \citealt{Bate:2012p24259})
and planet-planet scattering to wide orbits (\citealt{Boss:2006p18010}; \citealt{Scharf:2009p19732}; 
\citealt{Veras:2009p19654}).
Deciphering which mechanism dominates requires testable predictions from theory
and accurate demographics for PMCs, 
neither of which is currently available.

Observationally, the statistics of wide-separation PMCs are only beginning to be measured.
At projected separations of 400--4000~AU, \citet{Aller:2013bc} infer a frequency of 0.6~$\pm$~0.3\%
for substellar companions with masses of 15--60~\Mjup, which implies a frequency of $<$0.6\% for
companions with masses of $\sim$5--15~\Mjup \ based on their sensitivity limits ($\sim$5~\Mjup).  
This compares to an occurrence rate of 4$^{+5}_{-1}$\% by \citet{Ireland:2011p21592}
for 6--20~\Mjup \ companions between 200--500~AU.
Upper limits from high-contrast imaging surveys are yielding consistent results.  For example, \citet{Nielsen:2010p20955}
found that fewer than 20\% of solar-mass stars harbor $>$4~\Mjup \ planets between $\approx$20 and 500~AU 
at the 95\% confidence level.  At smaller separations of 10--150~AU, fewer than 6\% of stars harbor 1--20~\Mjup \ 
planets (\citealt{Biller:2013fu}).
PMCs therefore appear to be more common at several hundred AU
than several thousand AU, but larger samples are needed to make any strong
conclusions about their separation distributions.  

%broadly consistent with the distribution of 
%stellar companions to Sun-like stars (\citealt{Duchene:2013p25590}).

Regardless of their origin, wide-separation PMCs are valuable for
studying the atmospheres of young gas giants.  Their relatively large angular 
separations (typically $\sim$1--3$''$)
and modest contrasts ($\sim$5--7~mag in $H$-band) make them more 
accessible to spectroscopic observations than the current population of
``bona fide'' directly imaged planets, which generally reside at smaller separations and have 
less favorable contrasts ($>$10~mag).
Follow-up near-infrared spectroscopy of several PMCs has confirmed their low temperatures
 and youth (e.g., \citealt{Lafreniere:2008p14057};
\citealt{Bonnefoy:2010p20602}; \citealt{Bowler:2011p23014}; \citealt{Patience:2012p23718}).  
Overall, they exhibit several low-gravity features first identified in their young brown dwarf
counterparts in star-forming regions (e.g., \citealt{Lucas:2001p22099}; \citealt{Slesnick:2004jy}; 
\citealt{Allers:2007p66}; \citealt{Bonnefoy:2013p25217}) 
and the field (\citealt{McGovern:2004p21811}; \citealt{Kirkpatrick:2006p20500}; \citealt{Allers:2013p25314}), most notably 
angular $H$-band shapes and shallow $J$-band alkali line strengths.
Eventually, the near-IR spectra of wide PMCs can be used to map the influence of 
effective temperature and surface gravity (or equivalently, mass and age) 
on spectral morphology in order to calibrate
 more challenging spectroscopic studies of giant planets at smaller angular separations
 (\citealt{Bowler:2010p21344}; \citealt{Barman:2011p22098}; \citealt{Oppenheimer:2013p25039};
\citealt{Konopacky:2013p24810}).  

\citet{Kraus:2014hk} have recently discovered three new PMCs from follow-up imaging of 
candidate planetary companions reported in the literature.  Initially identified in 
binary surveys of the Taurus and $\rho$~Ophiucus star forming regions
 (\citealt{White:2001p21137}; \citealt{Ratzka:2005p22073}), 
 these PMCs have been shown to be comoving with their host stars at
 projected separations of hundreds of AU.  
 Here we present follow-up NIR spectroscopy for two of these new companions,
 FW~Tau~b and ROXs~42B~b.
  FW~Tau~b orbits the tight stellar binary FW~Tau~AB at a projected
  separation of 330 AU (2$\farcs$3).
 A member of the Taurus star forming region, the system has an age of $\sim$2~Myr, leading to
 a model-dependent mass of 5--13~\Mjup \ for the companion.
ROXs~42B~b has a similar mass estimate of 5--12~\Mjup \ and is bound to the tight
$\rho$ Oph binary ROXs~42B~AB at 140~AU (1$\farcs$2) separation.  The age estimate for the system
is $\sim$7~Myr based on the HR diagram positions of the stellar binary components
(\citealt{Kraus:2014hk}).  
Here we present spectroscopic follow-up observations of these
companions in an effort to confirm their cool temperatures and low masses.
Together these objects are among the lowest-mass
directly imaged companions known.

In addition, we present NIR spectroscopy of the PMCs GSC~6214-210~B and SR12~C (a.k.a., 2MASS~J16271951--2441403~C).
GSC~6214-210~B is a $\sim$14~\Mjup \ companion 
discovered by  \citet{Ireland:2011p21592} in a shallow imaging
search for PMCs in Upper Scorpius complex (5--10~Myr).
\citet{Bowler:2011p23014} presented $J$- and $H$-band spectroscopy of
GSC~6214-210~B which revealed a spectral type of L0$^{+2}_{-1}$.  
Notably, they found extraordinarily strong Pa$\beta$
emission originating from the companion, indicating vigorous accretion from a circum-substellar disk.
Here we present a $K$-band spectrum of GSC~6214-210~B.
Finally, we have also obtained a NIR spectrum of the 12--15 \Mjup \ PMC SR12~C, 
discovered by \citet{Kuzuhara:2011p21922} as part of a search for substellar companions
to members of the $\rho$~Oph complex.  With a separation of $\sim$1100~AU, it is among
the widest-separation PMCs known.  Kuzuhara et al. 
presented low-resolution NIR spectroscopy of this object; 
here we present the first moderate-resolution NIR spectrum.

\section{Observations}{\label{sec:obs}}

\subsection{Keck/OSIRIS NIR Spectroscopy of FW~Tau~b, \\ ROXs~42B~b, and GSC~6214-210~B}{\label{sec:keckobs}}

We observed FW~Tau~b with the OH-Suppressing Infrared Imaging 
Spectrograph (OSIRIS; \citealt{Larkin:2006p5570}) on 2011 October 4 UT
with the Keck II 10-m telescope using natural guide star adaptive optics (NGSAO).  
The 50 mas plate scale was used for both the $Hbb$ (1.473--1.803~$\mu$m)
and $Kbb$ (1.965--2.381~$\mu$m) filters, 
resulting in a 0$\farcs$8 $\times$ 3$\farcs$2 field of view and a resolving
power ($R \equiv \lambda$/$\delta \lambda$) of $\sim$3800.
The long axis was oriented parallel to the star-companion position angle and
the telescope was nodded along the detector by 1$''$ in an ABBA pattern.  
Sky frames were taken before
and after the observations.
Seeing measurements from the DIMM on CFHT were between 
$\approx$0$\farcs$8--1$\farcs$0 during the observations
 and modest levels of cirrus were present throughout the night.
We obtained six 300-s exposures in $Hbb$ and 13 300-s exposures in $Kbb$
at an airmass of 1.01--1.04.  Immediately after our science observations we
observed the A0V star HD~35036 at a similar airmass to correct
for telluric absorption.

In early 2012 OSIRIS moved to the Keck I telescope,
and on 2013 February 01 and 03 UT we obtained additional
NIR spectroscopy of FW~Tau~b with Keck~I/OSIRIS.
Our February 1 observations were carried out with
laser guide star AO using FW~Tau~AB as the on-axis tip-tilt star.
The wavefront sensor (WFS) $R$-band magnitude of the laser
was $\sim$9.7--10.2~mag.  Natural seeing was poor, with 
CFHT DIMM reporting $\approx$1$\farcs$2.
We obtained a total of 100~min of integration time in $Jbb$ and
60~min in $Kbb$ by nodding along the detector, which was
oriented perpendicular to the star-companion axis.
On February 3 we obtained a second epoch of $Hbb$ data
in NGSAO mode.  The WFS $R$-band magnitude of FW~Tau~AB 
was 14.0~mag, and seeing varied between $\approx$0$\farcs$5--1$\farcs$0.  
The observations were carried out in the same fashion
as the other bands, and we acquired a total of 30~min of 
integration time on the science target.
The same A0V standard, HD~35036, was used on both nights.

We obtained a $K$-band spectrum of ROXs~42B~b on 2011 August 20 UT
with OSIRIS on Keck~II using NGSAO.  CFHT-DIMM reported seeing 
measurements of $\approx$0$\farcs$6 during the observations.  In addition to
the \emph{bona fide} comoving companion ROXs~42B~b, \citet{Kraus:2014hk} also
identified a candidate companion located interior to ROXs~42B~b at a
mere 0$\farcs$55 from the primary star, but astrometry revealed that
this inner candidate is likely a background star.  At the time of our
OSIRIS observations, the status
of the inner object was more uncertain, so we attempted to obtain spectra of both of these 
companions simultaneously.  We chose the
35~mas plate scale to compensate between finely sampling the speckles
at small separations and obtaining a sufficiently large field of view (0$\farcs$56 $\times$ 2$\farcs$24)
to encompass both objects.  The orientation of the detector was parallel to
the inner companion-outer companion PA, and we nodded along the detector
axis in an ABBA pattern by 0$\farcs$4.  We obtained nine 300~s exposures
with the $Kbb$ filter at an airmass range of 1.43--1.57.
Prior to our science observations we observed the A0V standard HD 157734
at an airmass of 1.33 for telluric correction.

On 2012 June 2012 UT we observed GSC~6214-210~B with 
OSIRIS on Keck~I  using NGSAO.  
The long axis of the detector was oriented
perpendicular to the GSC~6214-210~A-B axis and was nodded
in an ABBA fashion by 1$''$.  We obtained eight 300-s exposures in $Kbb$
using the 50~mas plate scale.
The airmass range of our science observations was 1.36--1.46.
We then observed the A0V telluric standard HD~157170 at an airmass of 1.3.

Our OSIRIS observations of FW~Tau~b, ROXs~42B~b, and GSC~6214-210~B
were reduced in the same manner.  The OSIRIS Data Reduction Pipeline
was used for basic image reduction, which includes sky subtraction to remove
night-sky emission lines, corrections to account for DC bias shifts and electronic crosstalk
among the output channels, the removal of cosmic rays and bad pixels,
and rectification of the 2D spectra into 3D data cubes.  The rectification
matrices measured closest in time prior to the observation dates were used for
this last step.  Sky subtraction was performed with sky frames
obtained after the science observations 
for our 2011 FW~Tau~b data, whereas
consecutive ABBA nodded frames were used for ROXs~42B~b, GSC~6214-210~B,
and our 2013 FW~Tau~b data.

The science and standard spectra were extracted from the rectified data cubes using aperture photometry.
Aperture radii of 3 spaxels were centered on the  median-combined 
centroid position of the targets.  For our science frames we implemented sky subtraction 
measured from a large annulus surrounding the target.
Because the observations of the ROXs~42B companion were nodded by only 0$\farcs$4,
we used a smaller aperture radius of 2.5~spaxels without sky subtraction to avoid 
the negative dipole from the pairwise-subtracted nods.
Finally, telluric correction was performed for all our OSIRIS data using the \texttt{xtellcor\_general} 
routine from the Spextool IRTF data reduction package (\citealt{Cushing:2004p501}) based on the method 
described in \citet{Vacca:2003p497}.  

The inner point source located at 0$\farcs$55 from ROXs~42B is visible in our data and is clearly distinguishable
from nearby PSF speckles because of the wavelength-dependent PSF structure.
However this inner candidate is embedded in
the wing of the star's PSF, and there are several nearby bright speckles which make the extraction non-trivial.
We attempted to use spectral deconvolution with masking (\citealt{Thatte:2007p20429}), but because of
the close proximity of nearby speckles, the result is unreliable.  

\subsection{Gemini-North/NIFS NIR Spectroscopy of ROXs~42B~b}

We obtained $J$- and $H$-band spectroscopy of ROXs~42B~b with the 
Near-Infrared Integral Field Spectrometer (NIFS; \citealt{McGregor:2003p24390}) on the 
Gemini-North 8.1-m telescope 
in combination with the facility adaptive optics system ALTAIR (\citealt{Herriot:1998p24389}).
NIFS is an image-slicer integral field spectrograph with 0$\farcs$10~$\times$~0$\farcs$04
rectangular spaxels, providing imaging and $R$$\sim$5000 spectroscopy 
over a 3$\farcs$0~$\times$~3$\farcs$0 field of view.

Our $H$-band observations of ROXs~42B~b were obtained on 2012 May 09~UT.
$J$-band data were taken at two different dates, 2012 May 14~UT and 2012 Jul 04~UT,
 due to scheduling constraints.
The observations were carried out in queue mode (Program ID: GN-2012A-Q-48)
using NGSAO with the primary star
acting as the natural guide star ($R$ = 13.4~mag; \citealt{Zacharias:2013p24823}).
The Cassegrain rotator was set to track sidereal motion.
To avoid saturating the detector, 
we placed the primary off the array by 0$\farcs$70 
(the primary-companion separation is 1$\farcs$17) 
and oriented the instrument so that the
short axis of the spaxels were aligned with the primary-companion position angle
in order to better sample the primary star's PSF (Figure~\ref{fig:nifsimg}).
We obtained pairs of image cubes executed  
in an ABBA pattern with 1$\farcs$5 nods while reading the 
detector in ``Medium Object'' mode for our science observations.
Our $H$-band data ($H$ grating with the $JH$ filter)
consist of 27 300-s exposures (135 min total) over an airmass
range of 1.44--2.61; although this change in airmass is substantial, 
only eight frames were acquired above an airmass of 2.0.  
The May 2012 $J$-band data ($J$ grating with the $ZJ$ filter)
consist of 11 300-s exposures (55 min total) over an airmass
range of 1.49--1.73.
Our July 2012 sequence is longer and was better centered during transit;
28 300-s exposures (140 min total) were acquired
over an airmass range of 1.40--1.73.
After each science sequence a set of argon lamp frames were acquired at the
same sky position for wavelength calibration.  Standard A0V stars 
were observed in a dithered sequence immediately prior to or after the science sequences (Table~\ref{tab:obs}).
The airmasses of the standards are 2.2 for the 2012 May 09 data set,
1.6 for 2012 May 14, and 1.5 for 2012 July 04.
Flat fields, NIFS ``Ronchi'' calibration flats, and dark frames were taken at the end of the night.

Basic data reduction was carried out with the Gemini/NIFS processing package 
written for the Image Reduction and Analysis Facility (IRAF).
This includes cleaning the data of cosmic rays and bad pixels, deriving a wavelength
solution using the arc lamp frames, pairwise subtracting
nodded frames to remove sky features, and rectifying the 2D spectra into 3D image cubes.
We modified the default NIFS scripts to assemble individual cubes for both 
the science and standard data without correcting them 
for telluric absorption.
ROXs~42B~b is clearly visible in the wing of the primary star's PSF (Figure~\ref{fig:nifsimg}, top panel).
To remove this contaminating flux, we fit the extended PSF halo with several analytic models and subtracted them from
the data while masking out the companion.
We used the curve fitting package \texttt{MPFIT} (\citealt{Markwardt:2009p14854}) to separately test three functions --- 
a Gaussian, Lorenztian, and Moffat (\citealt{Moffat:1969p24391}) ---
by fitting and subtracting them from each column at every wavelength channel for all of our 
$J$- and $H$-band science observations (Figure~\ref{fig:nifsimg}, bottom panel).
The spectra were then extracted with aperture radii of 3 spaxels (0$\farcs$12).
The resulting spectra are in agreement for the three different representations of the extended PSF,
indicating that systematics from the subtraction process are minimal.
Telluric correction was performed with the \texttt{xtellcor\_general} 
routine in the IRTF/Spextool data reduction package.
In Figure~\ref{fig:roxs42_jcomp} we compare our telluric-corrected NIFS $J$-band spectra of ROXs~42B~b from May and July 2012.
The lower S/N spectrum from May contains many lines that appear to be telluric in origin, and are probably
a result of poor sky subtraction of the data.  We therefore adopt the higher S/N July spectrum for this work,
although we note that the overall shape of the extracted spectra are very similar at both dates.

\subsection{IRTF/SpeX NIR Spectroscopy of SR~12~C and 2MASS 0437+2331}{\label{sec:obsirtf}}

We observed SR12~C with IRTF/SpeX in short cross-dispersed (SXD) mode on 2012~July~11~UT.
The slit was set to 0$\farcs$8, resulting in a spectral resolution of $\sim$750.  We obtained 16 nodded pairs 
of SR12~C (32 individual frames) taken in an ABBA pattern at an airmass of $\approx$1.4.  
Individual exposure times were 150~s, providing a total on-source time of 80 min.  
Light cirrus was present throughout the night,
and the DIMM mounted on CFHT reported seeing measurements of 0$\farcs$6--0$\farcs$7 
during the observations.  We observed the A0V star HD 145127 at a similar airmass for telluric correction.
The data were reduced with the IRTF Spextool data reduction pipeline.

As a comparison template for our spectra of ROXs~42B~b and FW~Tau~b, we also targeted 
the latest-type member of the Taurus star-forming region currently known,  
2MASS~J04373705+2331080 (herinafter 2M0437+2331; \citealt{Luhman:2009p24392}), with IRTF/SpeX.  
The L0 optical spectrum obtained by Luhman et al.
revealed strong H$\alpha$ emission and spectroscopic indications of low surface gravity, confirming 
the object's youth.
At the extremely young age of Taurus, the inferred mass of 2M0437+2331 is $\sim$4--7~\Mjup.
Our observations of  2M0437+2331 were made on 2012 January 22~UT in prism mode using a 0$\farcs$8
slit width ($R$~$\sim$~90).  We acquired a total of 20~min (10~$\times$~120s) of data while nodding the
telescope along the slit.   Immediately afterwards we targeted the A0V standard HD~27761.  The data were
reduced and extracted using Spextool.

\subsection{UH 2.2~m/SNIFS Optical Spectroscopy of PMC Host Stars and FW Tau b}

In addition to spectroscopy of new PMCs, we are also 
gathering a uniform set of optical spectra of their primary stars with the University of Hawai'i~2.2-m telescope's 
SuperNova Integral Field Spectrograph  (SNIFS; \citealt{Lantz:2004p22123})
to characterize the host stars in a homogeneous manner (e.g., \citealt{Bowler:2011p23014}).
SNIFS is an integral field unit with a resolving power of $\sim$1300 that simultaneously
samples  blue (3000--5200~\AA) and red (5200--9500) spectral regions.
We observed ROXs~42B~AB and FW~Tau~AB on 2012 May 20~UT and 2011 October 08~UT,
respectively.   
ROXs~42B~AB was targeted for 4~min at an airmass of 1.4. 
with outstanding seeing ($\sim$0$\farcs$5) as reported by the DIMM mounted on CFHT.  
We acquired 15~min of data for FW~Tau~AB at an airmass of 1.02. 
Light cirrus was present throughout the night and seeing was poor ($\sim$1$''$). 
The O/B-type stars HR~718 and HD~93521 were used as standards.

Similarly, we targeted the PMC host star 1RXS~J160929.1--210524~A with SNIFS on 2012~May~20~UT.
We obtained 147~s of integration time at an airmass of 1.3.  Seeing ranged from 0$\farcs$4 to 1$\farcs$0 
throughout the night.  
Note, however, that dome seeing at the UH~2.2~m telescope limits the angular resolution 
of our SNIFS observations to $\gtrsim$0$\farcs$8.
The white dwarf EG131 and B-type star HR~5501 were used as spectrophotometric standards.

Details about our SNIFS data reduction procedures can be found in 
\citet{Mann:2012p24200}.
To summarize, the data were processed with the SNIFS data reduction pipeline 
(\citealt{Aldering:2006p22159}; \citealt{Scalzo:2010p22162}), which
includes flat, bias, and dark corrections; rectification of the data into spectral cubes; 
wavelength calibration using arc lamps; cleaning the data of bad pixels and
cosmic rays; and spectral extraction with a PSF model.

Upon close inspection of our SNIFS observation of FW Tau AB, we found
that the companion FW Tau b was also visible in the data, but only
in discrete emission lines and not in continuum.
On 2013~Jan~02~UT we obtained deeper SNIFS observations
of the FW Tau system to increase the S/N of our detection.
We acquired a total of 1~hr of data (two 30-min cubes) at an airmass of 
1.01.  
The seeing was excellent, with CFHT's DIMM reporting values of 0$\farcs$4--0$\farcs$6.
We once again recovered the companion in emission; Figure~\ref{fig:snifsimg} shows
SNIFS images of the system on and off the H$\alpha$ line.
The PSF from FW Tau AB contaminates our spectrum of the companion,
so for both data sets (Oct 2011 and Jan 2013) we extracted the spectrum of FW~Tau~b
using aperture photometry and then fit and subtracted our spectrum of the primary 
to remove its contaminating flux.  (This implicitly assumes the spectra of FW~Tau~AB
and FW~Tau~b differ.)  The residuals reveal a rich spectrum of emission lines in
FW~Tau~b, most of which are not seen in the spectrum of FW~Tau~AB, 
but our data do not detect the continuum.

Our spectrum of FW~Tau~AB was flux-calibrated using standards
from each night; we estimate the uncertainty to be
$\sim$10\% based on observations of the same star targeted on six nights as part of a separate
program with SNIFS (\citealt{Mann:2013fv}; \citealt{Lepine:2013p25027}).  FW~Tau~AB is a variable star, 
so this approach is more accurate than using archival photometry for flux calibration.
The spectra of FW~Tau~b were then absolutely flux calibrated using the ratio of H$\alpha$ line strengths 
between FW~Tau~AB and b (20.7 for our October 2011 data, 15.9 for our January 2013 data).

 \subsection{Keck/NIRC2 Imaging of FW~Tau~ABb}
 
On 2013 Feb 4 UT we imaged FW~Tau~ABb with Keck II/NIRC2 using NGSAO to check for long-term variability
in the system.  Seeing was highly variable throughout the night, ranging from $\sim$0$\farcs$6--2$''$.  We obtained 
five images in the narrow camera mode (10$\farcs$2$\times$10$\farcs$2 field of view) with integration times 
of 8.5~seconds per frame with the $K'$ filter.  In all of the images FW~Tau~AB is easily resolved and FW~Tau~b is visible.
Each image was dark-subtracted and flat-fielded after removing bad pixels and cosmic rays.
Optical distortions were corrected using the narrow camera distortion solution created by B. Cameron (private communication).

Photometry and astrometry for FW~Tau~AB is calculated by fitting an analytic model composed of three elliptical Gaussians
to each component as described in \citet{Liu:2008p14548}.  We measure a flux ratio of $\Delta K'$=0.091~$\pm$~0.011~mag,
a separation of 97.3~$\pm$~2~mas, and position angle of 316.7~$\pm$~0.4$^{\circ}$, which represent the
mean and standard deviation from the five images.  We adopt the sky orientation on the detector and pixel scale 
from \citet{Yelda:2010p21662}.  Our data show significant outward orbital motion compared to the $HST$ astrometry of 
75~$\pm$~5~mas and 3.4~$\pm$~3$^{\circ}$ from  \citet{White:2001p21137}.

Relative photometry between FW~Tau~AB and b is computed in a similar fashion.  A model of three elliptical Gaussians is
fit to the tertiary, yielding a relative flux ratio of $\Delta K'$=6.3~$\pm$~0.4~mag between the A-B pair and the outer companion.  
Despite these large uncertainties caused by poor seeing conditions, our photometry is consistent with
that from \citet{Kraus:2014hk} at the 1-$\sigma$ level, indicating no substantial $K$-band variability between 2008 and 2013. 
We measure a separation of 2307~$\pm$~4~mas and position angle of 296.0~$\pm$~0.10$^{\circ}$ relative to FW~Tau~A,
in agreement with the astrometry from \citet[with an updated PA from \citealt{Kraus:2014hk}]{White:2001p21137}.

\section{Results}{\label{sec:res}}

\subsection{Spectral Types and Reddenings for the Primary Stars}{\label{sec:prispt}}

Accurate reddening measurements are needed to correct for possible 
extinction in the spectra of companions.  
We fit our (unresolved) SNIFS spectra of ROXs~42B~AB and FW~Tau~AB for spectral type and visual extinction ($A_V$)
using optical templates from \citet{Pickles:1998p17673} for K3--K7
types and mean-combined SDSS templates from \citet{Bochanski:2007p19461} for M0--M9 types.
The reddening curve of \citet{Fitzpatrick:1999p24397} was 
used with $A_V /E(B-V) = 3.1$.  The spectra were fit using $\chi^2$ minimization
between 6000--8500~\AA \ by reddening each template in steps of 0.1~mag in $A_V$, interpolating onto the wavelength
grid of our science spectrum, and scaling to the observed spectrum using the maximum likelihood method 
described in \citet{Bowler:2009p19621}.
The results are shown in Figures~\ref{fig:chi2primary} and \ref{fig:primaryspt};
as expected, there is a strong covariance between spectral type and extinction,
with earlier types being more strongly reddened than later types to match the same slope.
The best-fitting spectrum to ROXs~42B~AB is M1~$\pm$~1 with a modest reddening of $A_V$=1.7$^{+0.9}_{-1.2}$~mag.
Here we adopt an uncertainty of one subclass for the spectral type and the corresponding $A_V$ range for the 
extinction.  (Our errors are estimated ranges rather than 1-$\sigma$ confidence intervals.)  
Our best-fit spectral type is comparable to M0~$\pm$~1 from \citet{Bouvier:1992p24486}.
For FW~Tau~AB, we find a spectral type of M6~$\pm$~1 and $A_V$=0.4$^{+1.3}_{-0.4}$~mag;
this compares well with the measurement of M5.5~$\pm$~0.5 from \citet{Briceno:1998p24484} and the 
reddening of $A_V$=0.00--0.83~mag derived by \citet{White:2001p21137} for the system based on photometry.
FW~Tau~AB exhibits Balmer emission lines (H$\alpha$ through H10) and strong Ca H+K emission,
both signs of strong chromospheric activity associated with youth.

We use the same fitting scheme to revisit the spectral types and reddening values
of the primary stars GSC~6214-210~A and 1RXS~J1609--2105~A, whose PMCs we use
as comparative templates for ROXs~42B~b (Section~\ref{sec:roxs42bb}).  
Our SNIFS spectrum of GSC~6214-210~A was first published in \citet{Bowler:2011p23014}
while our spectrum of 1RXS~J1609--2105~A is presented here for the first time.
The results of the fits are shown in Figures~\ref{fig:chi2_gsc_rxs_primary_spt} and 
\ref{fig:primary_gsc_rxs_spt}.  The best match to GSC~6214-210~A between 6000--8500~\AA \
is a K5~$\pm$~1 template reddened by $A_V$=0.5$^{+0.2}_{-0.5}$~mag.  In \citet{Bowler:2011p23014},
we found a spectral type of K7~$\pm$~1 but did not account for possible extinction; our revised
classification and slight reddening is similar to that found by \citet{Bailey:2013p25305}.
1RXS~J1609--2105~A is an excellent match to the M0 template, and we adopt a spectral type
of M0~$\pm$~1 with a slight reddening of $A_V$=0.1$^{+0.3}_{-0.1}$~mag for the system.
This agrees with the M0~$\pm$~1 type found by \cite{Preibisch:1998p21909} and is slightly
later than the K7~$\pm$~1 classification made by \citet{Lafreniere:2008p14057} based on
high-resolution optical spectroscopy.

\subsection{PMC Near-IR Spectral Types}

Our new NIR spectra of GSC~6214-210~B, SR~12~C, ROXs~42B~b, and FW~Tau~b 
are presented in Figure~\ref{fig:pmc_specfig}.  Each bandpass for GSC~6214-210~B, 
ROXs~42B~b, and FW~Tau~b was flux calibrated to the photometry listed in 
\citet{Kraus:2014hk}, and the spectra were Gaussian smoothed 
from their native resolution to $R$$\approx$1000.  
Our spectrum of SR~12~C was obtained with IRTF/SpeX, which simultaneously samples the 
1--2.5~$\mu$m region and produces reliable spectral slopes consistent
with photometry (\citealt{Rayner:2009p19799}),
so no relative flux calibration was performed for that object.
We smoothed the spectrum of SR~12~C to $R$$\approx$500 to increase the signal-to-noise, although
it is still rather modest in $J$-band ($\sim$30).  

The de-reddened spectra are overplotted in gray in Figure~\ref{fig:pmc_specfig}, 
which assumes the same reddening that we found for the primary stars in Section~\ref{sec:prispt}.  
Lacking an optical spectrum for SR~12~AB, we use extinction values from the literature for 
that system.  A large range of estimates exist: \citet{Sartori:2003p24526} cite 
0.0~mag based on the $V$--$I_C$ color index of SR~12~AB; \citet{Marsh:2010p24525} derive
a value of 2.3~$\pm$~1.6~mag from fitting reddened stellar models to 1--4~$\mu$m photometry; 
and \citet{Wahhaj:2010p24524} infer a value of 7.1~mag based on the $J$--$K$ color excess.
\citet{Kuzuhara:2011p21922} find that de-reddening their spectrum of SR~12~C by $A_J$=0.5~$\pm$~0.2~mag
produces a good fit to an M9 template spectrum.  Using the empirical relation between
$A_V$ and $A_J$  from 
\cite{Rieke:1985p24581},
this translates into an optical extinction of $A_V$=1.7~mag.
This is similar to the value of \citet{Marsh:2010p24525}, who used a broad wavelength range
for their estimation of SR~12~AB, so we adopt $A_V$=1.7~mag for the SR~12~ABC system 
in this work.

Except for FW~Tau~b, which we discuss in Section~\ref{sec:fwtaub}, our spectra 
resemble previously known low-gravity late-M/early-L type objects with shallow $J$-band alkali
absorption features and angular $H$-band shapes (\citealt{Lucas:2001p22099}; 
\citealt{Gorlova:2003p19016}; \citealt{McGovern:2004p21811};  \citealt{Allers:2007p66}).
Below we describe each spectrum individually and compare them to the new 
spectral classification system for young late-M and L dwarfs from \citet{Allers:2013p25314}.

\subsubsection{GSC~6214-210~B: M9.5~$\pm$~1.0}

\citet{Allers:2013p25314} use a range of spectral indices
to link NIR spectral types to optical classification schemes (e.g., \citealt{Cruz:2009p19453}) 
for young late-M and L dwarfs.  
They found that comparing the NIR spectra of young objects to standards in the field
leads to NIR types that are later than their optical counterparts by $\sim$1 subtype, a phenomenon
also  previously noted for late-M dwarfs (e.g., \citealt{Luhman:2004p22101}; \citealt{Herczeg:2009p18433}).  
In \citet{Bowler:2011p23014}, we compared the  $J$- and $H$-band spectra of GSC~6214-210~B
to M- and L-type templates in the field from \citet{Cushing:2005p288} and \citet{Rayner:2009p19799}, and the 
Upper Scorpius star forming region from \citet{Lodieu:2008p8698}. However, the Lodieu et al. study 
also made use of field templates for part of their classifications, possibly biasing our original classification.
Moreover, follow-up spectroscopy of these Upper Sco brown dwarfs has
revealed that many appear to be up to $\sim$2 subclasses earlier at optical wavelengths than 
the NIR estimates (\citealt{Herczeg:2009p18433}; 
\citealt{Biller:2011p22107}; \citealt{Lodieu:2011p24565}; \citealt{Bonnefoy:2013p25217}).
Here we reassess the L0~$\pm$~1 NIR spectral type we found for GSC~6214-210~B 
using the new Allers \& Liu classification sequence, our updated reddening measurement for the
primary star, and our new $K$-band spectrum.

The left-most panel of Figure~\ref{fig:gsc6214_vlgcomp} compares the 
flux-calibrated and de-reddened 1.2--2.4~$\mu$m spectrum 
of GSC~6214-210~B to the  ``very low-gravity'' spectral sequence of \citet{Allers:2013p25314}.
Normalized to the 1.66--1.68~$\mu$m region, the SED of GSC~6214-210~B is somewhat
redder than the young M8 template but bluer than the L0 template.  The M9 spectrum is a reasonable
match at $J$ and $H$, but the $K$ band of GSC~6214-210~B is redder than the template.
The right three panels of Figure~\ref{fig:gsc6214_vlgcomp} show similar normalized comparisons
for the individual bands.  GSC~6214-210~B is a good fit to the M9 template in $J$ band,
the M8 template in $H$-band, and L1--L2 for $K$ band.
The results are summarized in Table~\ref{tab:nirspt}.
Altogether we average the $JHK$ results and adopt a NIR spectral type of M9.5~$\pm$~1.0.
Interestingly, our $H$-band spectrum of GSC~62140219~B is significantly broader than
the other low-gravity templates in Figure~\ref{fig:gsc6214_vlgcomp}.  This may point to 
a somewhat higher surface gravity and hence older age for the system, though 
the relative change in $H$-band continuum shape does not uniquely track with changes in age (\citealt{Allers:2013p25314}).
We also note that, despite the strong Pa$\beta$ emission detected in $J$ band in July 2010, we do
not observe similar accretion-induced emission from Br$\gamma$ in $K$ band in June 2012.
\citet{Natta:2004p22062} found that only 2/8 young brown dwarfs showing Pa$\beta$
emission also showed Br$\gamma$ emission, so this is not unusual for substellar objects.

\subsubsection{SR~12~C: M9.0~$\pm$~0.5}

Our spectrum of SR~12~C (Figure~\ref{fig:pmc_specfig}) exhibits an angular $H$-band shape, 
an indicator of very low gravity (youth)  for low-temperature objects.  
The $J$-band region is relatively featureless; we do not detect the \ion{K}{1} alkali lines at 
1.243 and 1.252~$\mu$m, which 
may be due to the modest S/N in that region.  Strong steam absorption is evident at 1.35~$\mu$m.
Figure~\ref{fig:sr12c_vlgcomp} compares the SED and individual bands
of SR~12~C to the \citet{Allers:2013p25314} sequence.  Our de-reddened spectrum
of SR~12~C is an
excellent match to the M9 spectrum from 1.1--2.4~$\mu$m.  The best matches
to the $J$, $H$, and $K$ bands are the M8, M9, and M9--L0 templates, 
respectively (Table~\ref{tab:nirspt}).  We adopt a NIR classification of M9.0~$\pm$~0.5, which agrees
with the analysis of \citet{Kuzuhara:2011p21922} for their low-resolution
NIR spectrum.

\subsubsection{ROXs~42B~b: L1.0~$\pm$~1.0}{\label{sec:roxs42bb}}

The most prominent feature in our spectrum of ROXs~42B~b is the strong angular morphology of the $H$-band, which
is especially steep on the blue side compared to other low-gravity substellar objects
like SR~12~C.  The $J$-band region reveals strong FeH absorption at 1.19--1.20~$\mu$m and relatively
shallow \ion{K}{1} lines at 1.243 and 1.252~$\mu$m.
Compared to the Allers \& Liu sequence (Figure~\ref{fig:roxs42_vlgcomp}), the SED of
ROXs~42B~b resembles L1--L2 templates.  Individual band comparisons are best matched
with M9--L2 templates (see Table~\ref{tab:nirspt}).  
While the red end of the $H$ band matches M8--L2 objects
equally well, the blue end best matches the L1 and L2 templates.  Likewise, the depth of the 
steam absorption near 2~$\mu$m is similar to the L1 and L2 standards.  
The $J$ band is bluer than the L2 template but slightly redder than the M9 standard.  We emphasize
that our comparison in $J$ band suffers the most uncertainty since the reddening 
of the system, $A_V$=1.7$^{+0.9}_{-1.2}$~mag, is only modestly constrained from 
fitting the optical spectrum of the primary.  While this uncertainty has little
impact on the $H$ and $K$ bands, it has a noticeable effect on the $J$-band classification.
For $A_V$=2.6~mag, the best-fitting $J$-band type is M9, whereas for  $A_V$=0.5~mag, the best match is L0--L1.
Altogether, we assign ROXs~42B~b a near-infrared spectral type of L1.0~$\pm$~1.0.

Figure~\ref{fig:roxs42bb_comp} provides a detailed comparison of ROXs~42B~b with late-type substellar members
of other young associations.   \citet{Lodieu:2008p8698} identified 
USco~J160603.75--221930.0 (hereinafter USco~1606--2219) 
as a low-mass brown dwarf member of the Upper Scorpius star-forming region and  
inferred a spectral type of L2 from moderate-resolution 
near-IR spectroscopy.  
Optical spectroscopy by \citet{Herczeg:2009p18433} revealed an earlier optical type of M8.75
along with evidence of accretion from Balmer continuum excess and extraordinarily
strong H$\alpha$ emission ($\sim$750~$\pm$~80~\AA).  
ROXs~42B~b appears to be slightly later than USco~1606--2219
compared to the near-IR spectrum
from Lodieu et al., especially in $H$ and $K$ bands, 
assuming an extinction of $A_V$=0.0~mag for
USco~1606--2219 (\citealt{Herczeg:2009p18433}).

ROXs~42B~b is a good match in $H$ and $K$ to our prism spectrum of the young Taurus L0 
 object 2M0437+2331 (Figure \ref{fig:roxs42bb_comp} and \ref{fig:roxs42bb_2m0437_comp}).  The $J$-band spectrum of ROXs42B~b is 
significantly bluer than 2M0437+2331 assuming $A_V$=1.7~mag.  On the other hand, the spectra are in better agreement
if we use the lower bound of the extinction to the system ($A_V$=0.5~mag).
Here we have assumed zero reddening to 2M0437+2331, as found by \citet{Luhman:2009p24392}.
Although \citet{Oliveira:2013p24566} recently suggest a larger reddening value of $A_V$=2.1--3.3~mag
to 2M0437+2331 based on NIR spectral type matching, we adopt the value from Luhman et al. 
for this work since it is derived from optical data, which is more sensitive to reddening.

We also compare ROXs~42B~b with the young L-type companions AB~Pic~b (\citealt{Chauvin:2005p19642};
\citealt{Bonnefoy:2010p20602}) and 
1RXS~J160929.1--210524~b (herinafter 1RXS1609--2105 b; 
\citealt{Lafreniere:2008p14057}; \citealt{Lafreniere:2010p20986}).  
%Assuming 1.7~mag of optical extinction for ROXs42B~b,
Our de-reddened $J$-band spectrum of ROXs~42B~b is slightly bluer than that of the L0 companion AB~Pic~b
(recently reclassified by \citealt{Allers:2013p25314} and \citealt{Bonnefoy:2013p25217}),
but the $H$ and $K$ regions appear somewhat redder.  Compared with the L4 object 1RXS1609--2105 b,
ROXs~42B~b is a good match at $H$ and $K$, but 1RXS1609--2105 b is significantly bluer in $J$.
Altogether, our NIR classification of L1.0~$\pm$~1.0 for ROXs~42B~b broadly agrees with the
previous classification of these young brown dwarfs.  However, the shallow $J$-band slope relative
to 2M0437+2331 and AB~Pic~b may suggest that our reddening estimate of $A_V$=1.7~mag 
is too large.
 We also note that \citet{Currie:2014gp} recently presented a $K$-band spectrum of ROXs~42B~b and found a spectral
type of M8--L0, which is somewhat earlier than the type we infer from complete $JHK$ spectral coverage.

\subsection{Disk Accretion and Outflow Activity From FW~Tau~b}{\label{sec:fwtaub}}

\subsubsection{Overview}

Our spectra of FW~Tau~b reveal a wealth of permitted and forbidden emission lines
spanning the optical through NIR.  The optical IFU data from SNIFS are not sensitive
enough to detect continuum emission, but 
prominent emission lines from 4000---9000~\AA \ are visible.  
The NIR spectrum shows emission from molecular hydrogen and Pa$\beta$,  
but the continuum in $J$, $H$, and $K$ bands is relatively flat and featureless.
Continuum-subtracted H$_2$ images display slight, but significant, extended emission,
indicating spatially-resolved structure at small separations ($\lesssim$15~AU).  
These emission lines are commonly seen in Herbig-Haro (HH)
flows and their infrared counterparts, molecular hydrogen emission-line objects (MHOs).
Together with its extended morphology, this indicates that
FW~Tau~b is a young accreting object driving an outflow.

\subsubsection{Optical Emission-Line Spectrum}

Our flux-calibrated optical spectrum from January 2013 is shown in Figure~\ref{fig:fwtauemission}.
We detect forbidden emission lines from $[$\ion{S}{2}$]$ $\lambda \lambda$4069, 6716, 6731; 
$[$\ion{O}{1}$]$ $\lambda \lambda$5577, 6300, 6464; and $[$\ion{Fe}{2}$]$ $\lambda$8617,
as well as permitted \ion{Ca}{2} $\lambda \lambda$8498, 8542, 8662, and strong ($>$290~\AA) H$\alpha$ emission.  
A possible detection of H$\beta$ emission is visible, but at weak significance.  
In addition, we note an emission line at 8286.4~$\pm$~0.5~\AA \ that
is present at both epochs but which we are unable to identify\footnote{Two possibilities from
the National Institute of Standards and Technology Atomic Spectra Database 
(\citealt{Kramida:2012}, http://physics.nist.gov/asd) are \ion{Ne}{2} at 
$\lambda_\mathrm{air}$=8286.6 and \ion{Fe}{1} at $\lambda_\mathrm{air}$=8287.3.}.  
We do not reach the continuum level so we provide
lower limits on the equivalent widths together with line fluxes in Table~\ref{tab:emission}.
Line strengths are measured by fitting a Gaussian to each emission line
using the least-squares curve fitting routine \texttt{MPFIT} (\citealt{Markwardt:2009p14854}), except for the
partly blended $[$\ion{S}{2}$]$ $\lambda \lambda$6716, 6731, for which we simultaneously fit two Gaussians. 
Uncertainties in the line strength are derived in a Monte Carlo fashion by randomly adding noise 
drawn from a Gaussian distribution with a standard deviation equal to the 
spectral measurement uncertainty at that wavelength and then refitting the line profile.  The mean and
associated error from 200 trials are listed in Table~\ref{tab:emission}.
To compute lower limits on the equivalent widths, we adopt
continuum levels equal to the standard deviation of the data after
removing the emission lines (4.8$\times$10$^{-16}$ W m$^{-2}$ $\mu$m$^{-1}$ for our Oct 2011 data, 
1.1$\times$10$^{-16}$ W m$^{-2}$ $\mu$m$^{-1}$ for our Jan 2013 data).

Overall, the forbidden emission lines from FW~Tau~b resemble shock-excited gas typically 
seen in HH flows (\citealt{Schwartz:1983p24485}; \citealt{Reipurth:2001p51})\footnote{We follow the
definition of a Herbig-Haro object put forth by Reipurth (1999) for his catalog of Herbig-Haro flows:
``...flows that extend more than an arcsecond from a star, and which could in principle be studied
morphologically using proper coronagraphic techniques, are included in this catalog.  Stars with HH
emission lines but no appreciable extent have not been included."  Therefore, we do not classify FW~Tau~b as an HH object
since we do not detect extended emission in our (seeing-limited) SNIFS data.}.
Of particular interest is the $[$\ion{Fe}{2}$]$ $\lambda$8617 line, which is 
rarer in HH objects and their sources than $[$\ion{S}{2}$]$ and $[$\ion{O}{1}$]$ lines
(\citealt{Reipurth:2001p51}).
$[$\ion{Fe}{2}$]$ $\lambda$8617 emission is generally attributed to collisional excitation from shocks;
examples include the HH~34 energy source (\citealt{Reipurth:1986p24393}),
the HN~Tau (\citealt{Hartigan:2004p24518}) and HH 1 (\citealt{Nisini:2005p18946}) jets, 
and somewhat more enigmatic young stars such as V1647~Ori (\citealt{Aspin:2008p14717}), 
MWC~778 (\citealt{Herbig:2008p11354}), and PTF~10nvg (\citealt{Covey:2011ks}; \citealt{Hillenbrand:2013gy}).

\citet{Raga:1996p24623} find that HH objects can be divided into low-, intermediate-, and high-excitation
classes based on the $[$\ion{O}{3}$]$ $\lambda$5007/H$\beta$ and 
$[$\ion{S}{2}$]$~$\lambda \lambda$(6716+6731)/H$\alpha$ line ratios.
While we do not detect the $[$\ion{O}{3}$]$ $\lambda$5007 line, our 
$[$\ion{S}{2}$]$~$\lambda \lambda$(6716+6731)/H$\alpha$ 
measurement of 0.36 is comparable to intermediate- to high-excitation
HH objects.  (Our value is a lower limit, however, since activity and especially accretion from FW~Tau~b are also 
likely contributing to the H$\alpha$ emission).  
Compared to the sample of HH objects studied by Raga et al.,
our measurement of the electron density ($\sim 6\times10^3$ cm$^{-3}$; 
see Section~\ref{sec:physprop}) for FW~Tau~b is most
consistent with the high-excitation sources, which also tend to have large (10$^3$--10$^4$~cm$^{-3}$) values.

The \ion{Ca}{2} infrared triplet $\lambda \lambda$8498, 8542, 8662 has been observed in 
many young substellar objects (\citealt{Luhman:2003p22048}; \citealt{Scholz:2006p24628}).
\citet{Mohanty:2005p269} argue that for very low-mass objects, the presence of the \ion{Ca}{2}~$\lambda$8662 
emission line is an strong indicator of accretion.
In general, the \ion{Ca}{2} infrared triplet 
traces chromospheric activity in low-mass stars (e.g., \citealt{Busa:2007p24627}), 
but also appears in accreting stars (e.g., \citealt{Muzerolle:1998p33}; \citealt{Fernandez:2001p24625}) 
and some HH flows (\citealt{Reipurth:1986p24393}).
The relative strengths of the lines can be used to probe the optical depth of the emitting region,
which in principle can be used to distinguish between low-densiy outflow or high-density infall material.
Optically thin material yields line strengths of 1:9:5, whereas near-equal line strengths
imply an optically thick region --- a common observation in young accreting stars 
(e.g., \citealt{McGregor:1984p24626}; \citealt{Hamann:1989p24624}).
We measure ratios of 1:0.54:0.60, implying an optically thick origin in the form of accretion flows.

Unusually strong H$\alpha$ emission ($EW$=--454~$\pm$~224~\AA) was first detected from   
FW~Tau~b by \citet{White:2001p21137}
using narrow-band imaging with $HST$ in 1997.  They considered, but rejected, the possibility 
that the emission source was an HH knot since its appearance was point-like, the
neighboring weak-lined TTS FW~Tau~AB would have been an unlikely origin, 
and the object was also detected in continuum emission with the $F814W$ filter.
Figure~\ref{fig:fwtaub_specsed} shows the broadband detections and upper limits 
from \citet{White:2001p21137} compared with our SNIFS continuum upper limit.
Considering its strong H$\alpha$ emission at two epochs spanning $\sim$15 years, 
FW~Tau~b's high accretion activity appears to be long term and is worthy of continuous monitoring.
We note that there is a hint of the neighboring $[$\ion{N}{2}$]$~$\lambda$6583 line
in our data, but, if present, it is too weak and too blended with the H$\alpha$ line to resolve.

\subsubsection{NIR Continuum Spectrum}

Our OSIRIS spectra of FW~Tau~b are displayed in Figure~\ref{fig:pmc_specfig}.
As described in Section~\ref{sec:keckobs}, we obtained two epochs of NIR spectroscopy with OSIRIS:
$HK$ on 2011 Oct 04 and $JHK$ on 2013 Feb 01--03.
FW~Tau~AB is variable in the optical (hence its name), so care must
be taken with flux calibration since we do not have simultaneous
absolutely calibrated NIR photometry.
Indeed, archival sky surveys in the Vizier database
indicate $R$-band magnitude variations of $\approx$1~mag for FW~Tau~AB.
Photometric variability from stellar flares and star spots is minimal in the 
near-infrared (e.g., \citealt{Tofflemire:2011p24696}; \citealt{Davenport:2012p24694}; \citealt{Goulding:2012p24695}).
Accretion flows, hot spots, extinction variations, and disk heating can result in strong NIR variability
(\citealt{Skrutskie:1996p24697}; \citealt{Carpenter:2001p24700}; \citealt{Eiroa:2002p24701}; \citealt{Rice:2012p24698}),
but these effects are irrelevant for the Class III primary 
FW~Tau~AB (\citealt{Rebull:2010p24702}).
The lack of variability of the primary and companion is supported by
two epochs of $K$-band relative photometry of FW~Tau~b
by \citet{Kraus:2014hk}, which was
separated by about two years and showed no changes at the 0.1~mag level.
We therefore use the apparent magnitudes of FW~Tau~b in Kraus et al.
to absolutely flux calibrate our spectra.

Overall, the detailed shape of FW~Tau~b's NIR spectra do not resemble that of
a young late-type object as we expected from its photometry and age alone, 
which is similar to the Taurus L0 object 2M0437+2331.  For example, the $H$-band magnitude
and $J$--$K$ color of 2M0437+2331 is 16.13~$\pm$~0.15~mag and 1.94~$\pm$~0.33~mag 
(2MASS system; \citealt{Skrutskie:2006p589}), respectively, 
compared with 16.25~$\pm$~0.07~mag and 2.02~$\pm$~0.10 for FW~Tau~b (MKO system; \citealt{Kraus:2014hk}).
The flux calibrated spectra of both objects are shown in Figure~\ref{fig:fwtaub_specsed} for comparison;
despite their similar colors, their spectra are quite different.
Based on the evolutionary models of \citet{Baraffe:2003p587} and assuming
an age of 1--5~Myr, the FW~Tau~b's expected 
temperature is $\sim$2000--2100~K.  For field objects, this corresponds to
spectral types of $\sim$L1--L3 (\citealt{Golimowski:2004p15693}).
Instead, the $K$-band spectrum is mostly flat, with a slight decreasing slope
beyond 2.3~$\mu$m in our 2013 data and an apparently stronger feature in our 2011 data
between 2.30--2.35~$\mu$m.  The S/N of the smoothed spectra ($R$=1000) 
in Figure~\ref{fig:pmc_specfig}
are between 80--120 for our 2013 data and 60--90 for our 2011 data, which would easily
reveal CO absorption if present.  

The $H$ band spectra are of somewhat lower
quality, with smoothed S/N values of $\sim$30--50.  They appear
triangular in morphology, with the peak at a similar location near 1.7~$\mu$m, 
but with much shallower continuum slopes than for other young early-L 
dwarfs.  
Our $J$-band data have the lowest S/N (25--30 in our smoothed data, 
$\sim$8 in our raw data), so the spectral shape should be treated with caution.
Nevertheless, the data show a decline beyond 1.3~$\mu$m as is expected from
steam absorption in a cool atmosphere, but the rising flux below 1.25~$\mu$m is unusual
and may be a result of poor S/N.

\begin{center}
\emph{Near-IR Veiling?}
\end{center}

One explanation for the weak photospheric features 
is that they have been filled in (``veiled'') by excess continuum and/or
line emission from accretion activity or a warm inner disk.  This phenomenon was first
recognized in the optical spectra of accreting T Tauri stars as an additional
warm continuum component superimposed on an otherwise typical stellar spectrum,
resulting in weaker-than-expected absorption lines
(e.g., \citealt{Basri:1989p24709}; \citealt{Basri:1990p24705}; 
\citealt{Hartigan:1990p24711}; \citealt{Batalha:1993p24707}).
In the UV/optical, the excess emission is thought to originate from
shocks on the stellar surface caused by magnetospheric accretion (e.g., \citealt{Calvet:1998p13}).
A similar phenomenon occurs in the infrared as a result of reprocessed
light from an inner disk, adding continuum emission that can weaken or
completely mask otherwise strong photospheric lines (\citealt{Luhman:1998p24726};
\citealt{Luhman:1999p21747};  \citealt{Folha:1999p22272}; \citealt{JohnsKrull:2001p24704}; \citealt{McClure:2013hj}).
In addition to continuum excess, emission features from a warm ($>$1000~K) inner disk or stellar winds can 
also contribute to the diminished infrared absorption lines, the most notable being the first 
overtone bands of CO at 2.3~$\mu$m 
(\citealt{Carr:1993p24721}; \citealt{Casali:1996p24716}; \citealt{Najita:1996p24727}; 
\citealt{Fernandez:2001p24625}; \citealt{Berthoud:2007p24722}).  

Is veiling from a warm disk or accretion flow plausible for FW~Tau~b?
Based on thermal photometry of young ($\lesssim$5~Myr) brown dwarfs,
\citet{Liu:2003p20362} found that the inner regions ($\approx$2$R_*$) of brown dwarf disks
can reach temperatures of $\sim$1200~K, similar to those for T~Tauri stars.
This manifests as excess emission beyond $\sim$3~$\mu$m and, for particularly hot
disks, at $K$ band 
(e.g., \citealt{Luhman:1999p21746}; \citealt{Muench:2001jo}; \citealt{Liu:2003p20362}).
Moreover, brown dwarfs with especially high accretion rates (d$M$/d$t$ $\gtrsim$ 10$^{-9}$~M$_{\odot}$ yr$^{-1}$) 
can have accretion hot spots that dominate over the photospheric flux (\citealt{Mayne:2010p22280}).
This suggests that despite their cool temperatures, brown dwarfs can still 
experience veiling similar to their higher-mass counterparts.  
For FW~Tau~b, we measure a mass accretion rate of log(\emph{\.{M}})=--11.0~$\pm$~1.3~M$_{\odot}$ yr$^{-1}$
(Section~\ref{sec:accretion}); although this is below the extreme accretion rates required to dominate
the near-IR SED, it might produce partial veiling apparent in our spectra.

In a study of 110
Class I protostars, \citet{Connelley:2010p24432} found a correlation
between CO emission, high veiling (defined as veiling to the point where
no photospheric absorption features are detected), and Br$\gamma$ emission, 
which is probably caused by large mass accretion events 
(e.g., \citealt{Calvet:1991p15216}).
For FW~Tau~b, 
we do not detect Br$\gamma$ emission,
which is without exception observed by Connelley et al. when
CO is seen in emission.
Nevertheless, the Pa$\beta$ and
strong H$\alpha$ emission indicate that FW~Tau~b is actively accreting, 
supporting the idea that strong veiling could be a source of the weak near-IR features.
In general, near-IR veiling is characteristic of Class I objects
(e.g., \citealt{Greene:1996p20644}) and HH sources 
(\citealt{Reipurth:1997p24713}; \citealt{Davis:2011p22197}),
 pointing to a young evolutionary stage for FW~Tau~b.  

\begin{center}
\emph{An Edge-on Disk?}
\end{center}

Another interpretation is that FW~Tau~b is a low-mass star or brown dwarf
with an edge-on disk.  An earlier spectral type would help explain the lack of low-temperature 
photospheric features, and a nearly edge-on disk orientation 
could explain its faint optical and near-IR brightness.

Several edge-on disks around T Tauri stars and young brown dwarfs are known.
Their atypical orientations produce distinct observational signatures compared to 
more common inclinations.
Deep optical and NIR imaging often reveals dark, heavily extincted 
lanes intersecting an extended disk seen in scattered light 
(e.g., HV~Tau~C: \citealt{Monin:2000wi}, \citealt{Stapelfeldt:2003uw}; 
LkH$\alpha$~263C: \citealt{Jayawardhana:2002p19741}, \citealt{Chauvin:2002kn}). 
Although this morphology is not universal (\citealt{Huelamo:2010ic}), it is the most telling signature 
of an edge-on configuration.
The SEDs of edge-on disks generally exhibit double-peaked shapes 
with a broad trough at $\sim$5--20~$\mu$m (\citealt{Duchene:2010p22742}) and, at high disk inclinations,
 the 10~$\mu$m silicate feature is usually seen in absorption instead of emission
(\citealt{Watson:2004ch}; \citealt{Luhman:2007p20322}).

To test whether an earlier spectral type can plausibly fit our NIR spectrum of FW~Tau~b,
we created a simple model consisting of photospheric and blackbody emission from a single temperature disk.
We used the M0--M4 near-infrared spectra from \citet{Kirkpatrick:2010p21127} and 
the ``very low-gravity`` M5--L3 near-infrared standards from \citet{Allers:2013p25314} as
 photosphere templates.  
Each template was first flux-calibrated to FW~Tau~b's $J$-band magnitude, which will be minimally affected by warm
disk emission.  
Using the \texttt{AMOEBA} downhill simplex algorithm (\citealt{Press:2007p13558}), 
we fit a blackbody curve and photosphere
template to our 1.1--2.3~$\mu$m spectrum after applying $A_V$=0.4~mag of reddening 
to match that of FW~Tau~AB.  
In addition to the blackbody scale factor and temperature,
we allowed a third parameter of additional reddening to the system to vary to account for possible 
additional extinction from the disk itself.
Our spectrum of FW~Tau~b is well-reproduced with modestly-reddened (total $A_V$=0.80--1.2~mag)
late-M dwarf templates (M5--M8) with a hot ($T_{BB}$=1200--1330~K) disk.  
At later spectral types of L0--L3, the fit is somewhat worse 
($\chi^2_{\nu}$$\approx$5--7 compared to $\approx$2 for late-M types, where $\nu$=4268 degrees of freedom), 
but not implausible.
The bet-fit (M6, $T_{BB}$=1244~K, $A_V$=0.9~mag) is shown in Figure~\ref{fig:fwtaub_fitbb}.
The blackbody scale factor (=$R^2$/$d^2$) corresponds to an area of 
1.6~$R_{\odot}^2$ for the emitting region assuming a distance of 145~pc.
Although this simple model neglects other important factors like disk scattering and geometry, 
it shows that an underluminous late-M-type spectrum together with a warm disk might also
explain the properties of this companion.

In this edge-on disk scenario, the earlier spectral types of M5--M8 correspond to effective temperatures 
of $\sim$2700--3100~K based on the scale from \citet{Luhman:2003p22058}, which assumes
young M dwarfs have temperatures intermediate between field stars and giants.
At the system age of $\sim$2~Myr, these temperatures correspond to masses of $\sim$30~\Mjup \ to 0.15~\Msun \ 
based on the evolutionary models of \citet{Baraffe:1998p160}.

It is interesting to note that the few known mid- to late-M dwarfs with edge-on disks like 
TWA~30B (\citealt{Looper:2010p21597}) and 2MASS~J04381486+2611399 (\citealt{Luhman:2007p20322}) 
show deep photospheric absorption features superimposed on otherwise peculiar (and time varying for TWA~30B)
spectral energy distributions.   For example, both of these objects show pronounced 2.3~$\mu$m CO absorption, whereas 
FW~Tau~b has a flat $K$-band spectrum.  This may hint that both an edge-on disk \emph{and} veiling may
be present in FW~Tau~b. 

We can also examine the fraction of flux diminished by edge-on disks in these systems compared to
unobscured stars with the same age and spectral type.  TWA~30B is $\sim$5~mag fainter in $K$ band than its
companion, TWA~30A, despite having an earlier spectral type (M4.0~$\pm$~0.5 versus M5~$\pm$~1;
\citealt{Looper:2010p21597}).  This implies at least 5~mag of $K$-band extinction caused by its edge-on disk.
Compared to other M7.25 objects in Taurus from \citet{Luhman:2004p22196}, 2MASS~J04381486+2611399
is fainter in $K$ band by $\sim$1.5--3~mag.  Turning to FW~Tau~b, our best-fit spectral types in the
edge-on disk scenario are M5--M8.  This implies $\sim$2.5--5.5~mag of $K$-band extinction compared
with Taurus objects from \citet{Kraus:2006p20126} and \citet{Luhman:2004p22196}. 
The amount of extinction needed to explain the faintness of FW~Tau~b is therefore consistent with other
known edge-on disks around M stars, suggesting a similar disk geometry around FW~Tau~b is plausible.

\subsubsection{NIR Emission-Line Spectrum}

Several emission lines are visible in the NIR spectrum of FW~Tau~b (Figure~\ref{fig:fwtauemission}).  
In $K$ band, three low-excitation ($\nu$=1) H$_2$ rovibrational lines are
present at 2.034~$\mu$m, 2.122~$\mu$m, and 2.224~$\mu$m.
As described in detail in \citet{Beck:2008p22186}, several excitation mechanisms 
can give rise to low-$\nu$ H$_2$ transitions: shock-excited gas at temperatures 
of several thousand Kelvin (e.g., \citealt{CarattiOGaratti:2006p24570}),
UV-pumped fluorescence by Ly$\alpha$ emission (\citealt{Black:1987p24729}), 
or stellar heating of an inner disk by high-energy photons (\citealt{Maloney:1996p24731}).
The morphology of the H$_2$ emission is spatially extended (Section~\ref{sec:spatex}),
so at least some of it appears to be a result of shock-excited outflow activity. 
UV fluorescence is unlikely because of the absence of 
emission from higher H$_2$ transitions in our spectra 
(\citealt{Giannini:2002p24531}; \citealt{Giannini:2004p24558}).
Unfortunately, we cannot measure reddening and gas temperatures
using line ratios since we only detect $\nu$=1 lines; follow-up spectroscopy
with wavelength coverage beyond 2.4~$\mu$m would enable a more detailed 
analysis of the excitation conditions if the H$_2$ quadruple emission is present 
(e.g., \citealt{Gautier:1976p24732}; \citealt{Nisini:2005p18946}).

The $J$-band region shows emission from Pa$\beta$ and $[$\ion{Fe}{2}$]$ at 1.257~$\mu$m.
The $[$\ion{Fe}{2}$]$ emission originates from hot, dense gas at much a higher
temperature ($\sim$10$^4$~K) than the H$_2$ emission (\citealt{Hollenbach:1989p24734}; 
 \citealt{Nisini:2002p24546}; \citealt{Davis:2003p24537}; \citealt{Takami:2006p24513}), 
which dissociates at $\sim$4000~K (e.g., \citealt{Lepp:1983p24733}).  
Together these lines imply a stratification of gas temperatures and shock speeds 
at small spatial scales ($\lesssim$15~AU).  
We note that the 1.644~$\mu$m $[$\ion{Fe}{2}$]$ line may appear weakly in our 2013
data, but it is not strong enough to unambiguously identify above the pseudocontinuum.
If confirmed, the ratio of the 
1.257~$\mu$m and 1.644~$\mu$m lines can be used to probe extinction
to the emitting region (\citealt{Gredel:1994p24584}).

\subsubsection{Emission-Line Variability}{\label{sec:variability}}

Our multiple epochs of optical and NIR spectroscopy enable
emission line strength comparisons over a span of $\sim$15~months.
Figure~\ref{fig:snifs_var_fwtaub} shows our flux-calibrated spectra,
and $EW$ and line flux measurements are listed in Table~\ref{tab:emission}.
Most of the line fluxes agree.  The only ones showing significant 
($>$3~$\sigma$) variability are $[$\ion{S}{2}$]$~$\lambda$4067 and
H$_2$~1--0~S(2), suggesting some changes in the shocked regions
on $\sim$1~year timescales.

\subsubsection{Spatially Extended H$_2$ Emission}{\label{sec:spatex}}

Extended H$_2$ emission 
is commonly observed 
in the form of both bipolar jets and isolated outflows from young stars. 
NIR IFU observations in particular enable detailed spatial and kinematic
studies of molecular hydrogen outflows 
(e.g., \citealt{Beck:2008p22186}; \citealt{Davis:2011p22197}; \citealt{GarciaLopez:2013jv}).
To check for extended structure, we registered and coadded our nodded OSIRIS
cubes for each emission line seen in our 2011 and 2013 data,
subtracted continuum images adjacent to each line, and fit a 2D elliptical
Gaussian to the source.  The results of the fits are listed in Table~\ref{tab:fwspatfits}.
The FWHM of the Pa$\beta$, 1.257~$\mu$m $[$\ion{Fe}{2}$]$, and 2.223~$\mu$m 
H$_2$ 1--0 S(0) continuum-subtracted lines are consistent with the 
the neighboring continuum emission, meaning the emission is not resolved and these
lines originate from angular scales of $\lesssim$0$\farcs$05, or physical 
scales of $\lesssim$8~AU at a distance of $\sim$145~pc.  

On the other hand, 
the 2.122~$\mu$m  H$_2$ 1--0 S(1) and 
2.034~$\mu$m  H$_2$ 1--0 S(2) lines appear spatially extended compared to the continuum
emission.  This is verified by the FWHM measurements, which are about twice
as large for the continuum-subtracted images compared to the continuum region 
for our 2011 data.  Seeing and AO correction were substantially worse for our 2013 observations
($\sim$150~mas vs. 100~mas in $K$ band), but these data also reveal significantly extended
emission.  Figures~\ref{fig:fwtaub_h2_2011} and \ref{fig:fwtaub_h2_2013} show images of the emission lines,
adjacent continuum emission, and continuum-subtracted frames.   
The extended morphology appears asymmetric
in the south-west direction.  This is most evident for the 2011 1--0 S(1), 2011 1--0 S(2),
and 2013 1--0 S(2) lines, which have extensions at position angles of 240$^{\circ}$, +226$^{\circ}$, and +218$^{\circ}$,
respectively.  The extent of the lobe is $\sim$0$\farcs$1 from FW~Tau~b, or $\sim$15~AU.
Figure~\ref{fig:fwtaub_h2_profiles} shows cross sectional cuts of the 
H$_2$ 1--0 S(1) and 1--0 S(2) emission line images from 
Figures~\ref{fig:fwtaub_h2_2011} and \ref{fig:fwtaub_h2_2013} along their semi-major axes.  
Compared to normalized adjacent continuum emission, the
spatial extension and asymmetry are readily seen.
The line emission is kinematically unresolved so no velocity information can be retrieved.

Dozens of molecular hydrogen emission-line objects (MHOs) have been discovered
in Taurus (e.g., \citealt{Beck:2008p22186}; \citealt{Davis:2008p24593}; \citealt{Khanzadyan:2011p24813}).
Since the outflow from FW~Tau~b is resolved, it has been assigned the 
name MHO 750 as the next addition to the online catalog of MHOs 
(\citealt{Davis:2010p24599}\footnote{http://www.astro.ljmu.ac.uk/MHCat/.}; C. Davis, private communication, 2013), 
which follows IAU naming conventions.
None of the known MHOs in Taurus are obviously associated with the FW~Tau system.

\subsubsection{Physical Properties of the Emitting Gas}{\label{sec:physprop}}

The ratio of atomic and molecular emission lines spanning the optical to NIR 
can probe the physical conditions of 
the emitting region, including electron density, temperature, ionization fraction, and optical extinction 
(e.g., \citealt{Hartigan:1994p24602}).
Several reliable line diagnostics have been established.  
The $[$\ion{S}{2}$]$ $\lambda$6716 and $\lambda$6731 lines have similar excitation energies, 
but different collisional strengths and transition probabilities, so their ratio can be used to determine 
$n_e$ (\citealt{Osterbrock:1989p24612}).
The critical densities for $\lambda$6716 and $\lambda$6731 are
$\sim$1.5$\times$10$^{3}$~cm$^{-3}$ and $\sim$4$\times$10$^{3}$~cm$^{-3}$, respectively (\citealt{Appenzeller:1988p24579}).
At low densities ($n_e$$\lesssim$100~cm$^{-3}$), the line ratio is simply the ratio of statistical weights (1.5). 
At high densities ($n_e$$\gtrsim$10$^4$~cm$^{-3}$) collisional de-excitation becomes 
significant and forbidden lines get weaker, producing a line ratio (0.44) equal to the statistical weights modulated by
the transition probabilities .
We measure values of  0.53~$\pm$~0.08 and 0.7~$\pm$~0.3 for our 2013 and 2011 data, respectively,
implying an electron density of $\sim$6$^{+8}_{-3}\times10^3$~cm$^{-3}$.  (The large uncertainty
in $n_e$ is caused by the steep dependence of $\lambda$6716/$\lambda$6716 on $n_e$ near the
high-density limit.)
\citet{Raga:1996p24623} compute $n_e$ using the
$[$\ion{S}{2}$]$ $\lambda$6716/$\lambda$6731 line ratio for 45 HH condensations and find
that most have $n_e$ below 5$\times$10$^3$~cm$^{-3}$.
In comparison,
the shocked gas in the vicinity of FW~Tau~b is near the high end for HH objects.

The $[$\ion{Fe}{2}$]$ $\lambda$7155/$\lambda$8617 and 
$[$\ion{Fe}{2}$]$ $\lambda$12567/$\lambda$8617 line ratios can also be used to
probe electron density of the emitting region.
Unfortunately, the $\lambda$7155 line is not detected, 
but our upper limit for the ratio $\lambda$7155/$\lambda$8617 of $\lesssim$0.5 
implies electron densities of $\lesssim$10$^4$~cm$^{-3}$
(\citealt{Bautista:1996p24816}; \citealt{Bautista:1998p24815}). 
This is consistent with our measurement from the 
$[$\ion{S}{2}$]$ $\lambda$6716/$\lambda$6731 ratio.
On the other hand, both $\lambda$12567 and $\lambda$8617 are detected,
but the ratio we measure (0.74~$\pm$~0.24) suggests higher 
values of $>$10$^5$~cm$^{-3}$  (\citealt{Bautista:1998p24815}).  
This may indicate that multiple regions
with different physical conditions 
are producing the $[$\ion{Fe}{2}$]$ lines.

\subsubsection{Accretion}{\label{sec:accretion}

We use the empirical correlation 
between accretion luminosity ($L_\mathrm{acc}$) and  
Pa$\beta$ luminosity ($L_\mathrm{Pa \beta}$) from \citet[their Equation 2]{Natta:2004p22062}
to derive a mass accretion rate (\emph{\.{M}}).
A distance estimate of 145~$\pm$~10~pc together with our measured Pa$\beta$ line flux  of 3.1~$\pm$~0.4
erg~s$^{-1}$~cm$^{-2}$ yields a Pa$\beta$ line luminosity of log($L_\mathrm{Pa \beta}$/$L_{\odot}$)~=~--6.53~$\pm$~0.08~dex
and an accretion luminosity of
log($L_\mathrm{acc}$/$L_{\odot}$)~=~--4.9~$\pm$~1.3~dex.
The accretion rate and the accretion luminosity are related through
\emph{\.{M}}=$L_\mathrm{acc}$$R/GM$, where $R$ is object's radius and $M$ is its mass.

Since we do not have tight constraints on the mass of FW~Tau~b, we calculate accretion rates
corresponding to the two physical scenarios: a veiled planetary-mass companion, and an 
extincted low-mass star or brown dwarf with an edge-on disk.
In the former case, we adopt the mass estimate of 10~$\pm$~4~\Mjup \ from \citet{Kraus:2014hk}, which corresponds to a 
radius of 0.21$\pm$0.05~$R_{\odot}$ (\citealt{Baraffe:2003p587}) 
and a mass accretion rate of log(\emph{\.{M}})~=~--11.0~$\pm$~1.3~M$_{\odot}$ yr$^{-1}$.
This is consistent with accretion rate measurements of other substellar objects,
which roughly follow an \emph{\.{M}} $\propto$ $M^2$ relationship 
(\citealt{Natta:2004p22062}; \citealt{Mohanty:2005p269}; \citealt{Muzerolle:2005p55}; \citealt{Herczeg:2008p24117}).
On the other hand, the higher mass range of 0.03--0.15~\Msun \ corresponds to 
radii of 0.37--0.96~$R_{\odot}$ (\citealt{Baraffe:1998p160}) and a mass accretion rates of 
--11.4~$\pm$~1.3~M$_{\odot}$ yr$^{-1}$.

\section{Discussion and Conclusions}{\label{sec:disc}}

There is strong evidence that companions near the deuterium-burning
limit can form from multiple pathways.  The HR~8799 planets offer the best
support for non-hierarchical formation from a disk since initial constraints
on their orbits are consistent with coplanarity (\citealt{Esposito:2012hs}), 
and the planets are flanked by debris disks both inside and outside of their orbits (e.g., \citealt{Su:2009p20065}).
On the other hand, the tail of the initial mass function appears to continue down
to at least $\sim$5--10~\Mjup \ (e.g., \citealt{Luhman:2009p24392}; 
\citealt{Scholz:2012kv}; \citealt{Oliveira:2013p24566}; \citealt{Liu:2013gya}), 
and planetary-mass companions have been found orbiting low-mass brown dwarfs
with decidedly non-planetary mass ratios (\citealt{Chauvin:2004p19400}; 
\citealt{Navascues:2007dk}; \citealt{Bejar:2008p130}; \citealt{Todorov:2010p20562}; 
\citealt{Liu:2012p24074}); together these suggest that cloud fragmentation and/or
ejection can form objects that overlap with apparently bona fide planets (i.e., those born in a disk).
These dual origins pose problems for interpreting discoveries from
direct imaging planet searches, which are finding companions in this
ambiguous 10--15~\Mjup \ mass range (e.g., \citealt{Chauvin:2005p19642}; 
\citealt{Itoh:2005p19673}; \citealt{Delorme:2013p25184}; \citealt{Bowler:2013p25491}).

At the youngest ages, the disk properties of planetary-mass companions 
can provide constraints on their formation since dynamical scattering to 
wide orbits probably disrupts circum-planetary disks (\citealt{Bowler:2011p23014}; \citealt{Bailey:2013p25305}).  
\citet{Kraus:2014hk} find that five out of seven of the planetary-mass companions 
in their sample have a redder $K$--$L'$ color than field objects with the same spectral type,
indicating that disks may be common to wide low-mass companions at very young ages.
Strong Pa$\beta$ emission from GSC~6214-210~B unambiguously confirms the presence of an
accretion disk 
(\citealt{Bowler:2011p23014}), and we show here that the same is true for FW~Tau~b, suggesting that
it too probably formed \emph{in situ} at several hundred AU.  
Hydrodynamic simulations by \citet{Shabram:2013dm} show that 
planetary companions formed on wide orbits via disk instability have 
thick disks that are truncated at $\sim$1/3 of the companion's Hill radius.
GSC~6214-210~B and FW~Tau~b may be promising targets for follow-up observations
to directly compare with these predictions.

The remarkable outflow activity we identified from FW~Tau~b
provides some general clues about the formation and evolution of planetary-mass objects.
Jets and outflows are common products of the star formation process, 
helping to shed disk angular momentum, regulate stellar mass, and contribute
to turbulence in molecular clouds (e.g., \citealt{Lada:1985p24487}; 
\citealt{Richer:2000p24821}; \citealt{Quillen:2005p24820}).
In the substellar regime, disk accretion is commonly observed from young brown dwarfs, 
but evidence of outflows is much more rare because the mass accretion rate and mass loss
are orders of magnitude smaller (\citealt{Whelan:2005p24578};
\citealt{Whelan:2009p22203}).  
One of the best-studied substellar outflows 
is the $\sim$24~\Mjup \ TWA member 2MASS~J1207334--393254~A 
(\citealt{Whelan:2007p24592}; \citealt{Whelan:2012p24580}),
which until FW~Tau~b has been the lowest-mass jet-driving source known.  
The outflow itself appears collimated, bipolar, and altogether analogous to
those from low-mass stars. 
To date, a handful of outflows from brown dwarfs have
been detected in the optical through forbidden line emission (\citealt{Joergens:2012p24800}) 
and CO emission (\citealt{PhanBao:2008p24522}; \citealt{Monin:2013ep}).

The outflow from FW~Tau~b makes it among only a handful
of substellar objects showing such activity.  
If FW~Tau~b is confirmed to be planetary mass instead of a low-mass companion with an edge-on disk,
it will be the first planetary-mass object to drive an outflow.
That would suggest that the same physical processes governing accretion, magnetic collimation,
and angular momentum regulation  
 in stars also operate at the bottom of the IMF.  
Distinguishing the veiled planetary-mass scenario from the higher-mass
 edge-on disk hypothesis should therefore be a chief goal for future studies of this system.
One simple test is to search for weak photospheric features unique to M or L dwarfs in higher-S/N 
near-infrared spectroscopy; for example, the 2.21~$\mu$m \ion{Na}{1} doublet appears in
M dwarfs (though is weaker at lower gravities), but is absent in L dwarfs. 
Perhaps an outflow from such a low-mass companion should not be surprising since several 
free-floating planetary-mass objects are known to possess disks (e.g., \citealt{Luhman:2005p14774}), and 
 theoretical arguments suggest that outflows  
 might be natural byproducts of giant planet formation (\citealt{Quillen:1998p22297}).
Finally, we note that if the extended H$_2$ emission from FW~Tau~b is confirmed to be a jet, it will be the
first substellar molecular hydrogen outflow known.

  \acknowledgments
We are grateful to Bo Reipurth and Lynne Hillenbrand for helpful comments on the properties of FW~Tau~b,
Katelyn Allers for the low gravity spectral templates used in this work,
and Kimberly Aller and Will Best for carrying out some of the IRTF observations.
It is also a pleasure to thank Joel Aycock, Randy Campbell, Heather Hershley,
Jim Lyke, Julie Rivera, Hien Tran, Cindy Wilburn 
and the entire Keck staff for their support with the observations.
B.P.B. and M.C.L. have been supported by NASA grant NNX11AC31G and NSF grant AST09-09222.
Some of this work is 
based on observations obtained at the Gemini Observatory, which is operated 
by the Association of Universities for Research in Astronomy, Inc., under a 
cooperative agreement with the
NSF on behalf of the Gemini partnership: the National Science
Foundation (United States), the Science and Technology Facilities
Council (United Kingdom), the National Research Council (Canada),
CONICYT (Chile), the Australian Research Council (Australia),
Minist\'{e}rio da Ci\^{e}ncia, Tecnologia e Inova\c{c}\~{a}o (Brazil)
and Ministerio de Ciencia, Tecnolog\'{i}a e Innovaci\'{o}n Productiva
(Argentina)
We utilized data products from the Two Micron All Sky Survey, which is a joint project of the University of Massachusetts and the Infrared Processing and Analysis Center/California Institute of Technology, funded by the National Aeronautics and Space Administration and the National Science Foundation.
 NASA's Astrophysics Data System Bibliographic Services together with the VizieR catalogue access tool and SIMBAD database 
operated at CDS, Strasbourg, France, were invaluable resources for this work.
Finally, mahalo nui loa to the kama`\={a}ina of Hawai`i for their support of Keck and the Mauna Kea observatories.
We are grateful to conduct observations from this mountain.

\facility{{\it Facilities}: \facility{Keck:II (OSIRIS)}, \facility{Keck:I (OSIRIS)}, \facility{Keck:II (NIRC2)}, \facility{Gemini:Gillett (NIFS)}, 
\facility{IRTF (SpeX)}, \facility{UH:2.2m (SNIFS)}}

\newpage

%\bibliographystyle{apj}
%\bibliography{oct13}

\newpage

\clearpage
\begin{landscape}
\begin{deluxetable*}{lcccccccc}
%\hskip -.3in
\tabletypesize{\footnotesize}
%\rotate
\tablewidth{0pt}
\tablecolumns{9}
\tablecaption{Wide ($>$100~AU) Companions Under 15~\Mjup \ Around Stars\label{tab:pmc}}
\tablehead{
        \colhead{Object}     &   \colhead{Mass}     &  \colhead{Age}    &    \colhead{Separation}    &    \colhead{NIR SpT}  &  \colhead{Pri. Mult.\tablenotemark{a}} & \colhead{Pri. Mass}  & \colhead{Pri. SpT}  &  \colhead{References}                                              \\
        \colhead{}                &   \colhead{(\Mjup)}   &  \colhead{(Myr)}  &    \colhead{(AU)}               &    \colhead{}                  &  \colhead{}                           & \colhead{(\Msun)}             & \colhead{}    &  \colhead{} 
        }   
\startdata
WD~0806-661 B                  &      6--9                      &     2000~$\pm$~500      &    2500               &      Y?                       &   S  &   2.0~$\pm$~0.3                 &   DQ\tablenotemark{b}         &      1, 2, 3   \\
HD~106906 b                      &        5 $\pm$ 2          &      13~$\pm$~2            &       $\sim$650    &    L2.5~$\pm$~1    &  S  &  1.50~$\pm$~0.10               &   F5     &        4, 5  \\
1RXS~J1609--2105~b      &   8$^{+4}_{-2}$    &  $\sim$5                            &  $\sim$330          &  L4 $\pm$ 1            & S  & 0.85$^{+0.20}_{-0.10}$  & M0~$\pm$~1 &    6, 7, 8, 9   \\
ROXs~42B b                          &      6--14                    &   7$^{+3}_{-2}$               &  $\sim$140        &      L1 $\pm$ 1       & B &  0.89~$\pm$~0.08, 0.36~$\pm$~0.04 & M1 $\pm$ 1 &     9, 10          \\
FW Tau b                              &      6--14                    &   2$^{+1}_{-0.5}$           &   $\sim$330        &    pec                      & B & 0.28~$\pm$~0.05, 0.28~$\pm$~0.05  & M6~$\pm$~1 &      9, 10         \\
Ross 458~C                           &     6--12                      & 150--800                         &   1190                &     T8.5p $\pm$ 0.5   & B  & $\approx$0.6, $\approx$0.09  & M0.5~$\pm$~0.5, $\sim$M7  &      11, 12, 13, 14, 15, 16         \\
ROXs12 b                             &     12--20                      &  8$^{+4}_{-3}$                &  $\sim$210       &   $\cdots$              & S  & 0.87~$\pm$~0.08 & M0~$\pm$1  & 10, 17        \\
AB Pic B                                &      13--14                   &   $\sim$30                       & 250                    & L0 $\pm$ 1           & S  & 0.95~$\pm$~0.05\tablenotemark{c} & K2  &       18, 19         \\
CHXR 73 B                          &       13$^{+8}_{-6}$      &         $\sim$2                 &     $\sim$210      &   $\geq$M9.5     & S  & 0.30 &  M3.25~$\pm$~0.25  &      20, 21          \\
DH Tau B                              &        12$^{+10}_{-4}$  &         $\sim$1--2            &     $\sim$340     &  M9.25~$\pm$~0.25             & S  & 0.53 &  M2  &       17, 22, 23, 24        \\
GSC 6214-210 B                 &      14 $\pm$ 2          &         $\sim$5                  &     $\sim$320     &   M9.5 $\pm$ 1     & S  & 0.9~$\pm$~0.1 & K5 $\pm$ 1  &        8, 9, 25       \\
SR12 C                                 &         12--15              &         $\sim$2                  &      $\sim$1100   &    M9.0 $\pm$ 0.5  & B & 1.05~$\pm$~0.05, 0.5~$\pm$~0.1\tablenotemark{d}  &  K4~$\pm$~1, M2.5~$\pm$~1  &    9, 26    
\enddata
\tablenotetext{a}{Single (``S'') or binary (``B'').}
\tablenotetext{b}{WD~0806-661 is a DQ white dwarf.}
\tablenotetext{c}{Based on evolutionary models from \citet{Baraffe:1998p160} using the age (30~Myr, \citealt{Song:2003p25386}) and 
luminosity (0.5~$L_{\odot}$; \citealt{McDonald:2012cg}) of AB~Pic.}
\tablenotetext{d}{The mass of SR~12~A is inferred from the evolutionary models of \citet{Baraffe:1998p160} based on its age ($\sim$2~Myr) and 
luminosity (1.03~$L_{\odot}$; \citealt{Wahhaj:2010p24524}). Lacking a luminosity for SR~12~B, we use its effective temperature (3490~$\pm$~70~K, 
inferred from the spectral type-temperature conversion of \citealt{Luhman:1999p21746}) and age to compute a mass from evolutionary models.}

\tablerefs{
(1) \citet{Luhman:2011p22766}; 
(2) \citet{Luhman:2012p23487}; 
(3) \citet{Rodriguez:2011p22764}; 
(4) \citet{Bailey:2014et};
(5) \citet{Pecaut:2012p23696}; 
(6) \citet{Lafreniere:2008p14057};
(7) \citet{Lafreniere:2010p20986}; 
(8) \citet{Ireland:2011p21592}; 
(9) this work;
(10) \citet{Kraus:2014hk};
(11) \citet{Goldman:2010p22044};
(12) \citet{Scholz:2010p20993};
(13) \citet{Burgasser:2010p21472};
(14) \citet{Burningham:2011ed}; 
(15) \citet{Beuzit:2004p18015};
(16) \citet{Reid:1995p22125}; 
(17) \citet{Bouvier:1992p24486}; 
(18) \citet{Chauvin:2005p19642};
(19) \citet{Perryman:1997p534}; 
(20) \citet{Luhman:2006p19659};
(21) \citet{Luhman:2004hy}; 
(22) \citet{Itoh:2005p19673}; 
(23) \citet{White:2001p21137};
(24) \citet{Bonnefoy:2013p25217}; 
(25) \citet{Bowler:2011p23014}; 
(26) \citet{Kuzuhara:2011p21922}.
}

\end{deluxetable*}
\clearpage
\end{landscape}

\clearpage

\begin{deluxetable}{lcccccc}
\tabletypesize{\normalsize}
%\rotate
\tablewidth{0pt}
\tablecolumns{7}
\tablecaption{Spectroscopic Observations\label{tab:obs}}
\tablehead{
        \colhead{Object}               &   \colhead{Date}   &  \colhead{Telescope/} &    \colhead{Filter}             &    \colhead{Plate Scale}   & \colhead{Exp. Time}  &    \colhead{Standard}  \\
        \colhead{}   &   \colhead{(UT)}   &  \colhead{Instrument}   &    \colhead{}    &     \colhead{(mas)}  &  \colhead{(min)} &    \colhead{} 
        }   
\startdata

FW Tau b      &    04 Oct 2011  &   Keck-II/OSIRIS  &    $Hbb$   &   50  & 30  & HD 35036   \\
FW Tau b      &    04 Oct 2011  &   Keck-II/OSIRIS  &    $Kbb$    &   50  & 65 &   HD 35036   \\
FW Tau b      &    01 Feb 2013  &   Keck-I/OSIRIS  &    $Jbb$    &   50  & 100 &   HD 35036   \\
FW Tau b      &    01 Feb 2013   &   Keck-I/OSIRIS  &    $Kbb$    &   50  & 60 &   HD 35036   \\
FW Tau b      &    03 Feb 2013  &   Keck-I/OSIRIS  &    $Hbb$    &   50  & 30 &   HD 35036   \\
FW Tau AB+b      &    08 Oct 2011  &   UH~2.2~m/SNIFS  &    $\cdots$    &   400  & 15 &   HR~718, HD~93521   \\
FW Tau AB+b      &    02 Jan 2013  &   UH~2.2~m/SNIFS  &    $\cdots$    &   400  & 60 &   HR~718, HD~93521   \\
ROXs~42B b      &   09 May 2012   &   Gemini-N/NIFS  &    $H$\tablenotemark{a}  & 100$\times$40    & 135 &    HD 155379  \\
ROXs~42B b      &  14 May 2012    &   Gemini-N/NIFS  &    $J$\tablenotemark{b}    &  100$\times$40  & 55 &  HD 155379    \\
ROXs~42B b      &   04 Jul 2012   &   Gemini-N/NIFS  &    $J$\tablenotemark{b}      &  100$\times$40  &  140 &  HD 155379  \\
ROXs~42B b      &  20 Aug 2011  &   Keck-II/OSIRIS  &    $Kbb$   &   35     &  45  & HD 157734  \\
ROXs~42B AB+b   &    20 May 2012  &   UH~2.2~m/SNIFS  &    $\cdots$    &   400  & 4 &   EG131, HR~5501   \\
1RXS~J160929.1--210524~A   &  20 May 2012   &   UH~2.2~m/SNIFS  &    $\cdots$    &   400  &  2.5   & EG131, HR~5501  \\ 
GSC 6214-210 B      &    26 Jun 2012   &   Keck-I/OSIRIS  &    $Kbb$   &   50  & 40  &  HD 157170    \\
SR12 C      &   11 Jul 2012  &   IRTF/SpeX-SXD\tablenotemark{c}  &    $\cdots$    &   150  &   80  &   HD 145127  \\
2M0437+2331  & 22 Jan 2012  &   IRTF/SpeX-prism\tablenotemark{c}    &    $\cdots$    &   150  & 20   &  HD 27761 
 
\enddata
\tablenotetext{a}{NIFS $JH$ filter with $H$-band grating.}
\tablenotetext{b}{NIFS $ZJ$ filter with $J$-band grating.}
\tablenotetext{c}{A slit width of 0$\farcs$8 was used.}

\end{deluxetable}

%\clearpage

\begin{deluxetable}{lccccc}
\tabletypesize{\normalsize}
\tablewidth{0pt}
\tablecolumns{6}
\tablecaption{PMC Near-Infrared Spectral Types\label{tab:nirspt}}
\tablehead{
        \colhead{Object}               &   \colhead{1.1--2.4~$\micron$}   &  \colhead{$J$-Band}     &    \colhead{$H$-Band}   &    \colhead{$K$-Band}  &    \colhead{Adopted}  
         }   
\startdata
GSC 6214-210 B    &   M9   &    M9   &    M8   &   L1--L2   &  M9.5 $\pm$ 1   \\
SR12 C    &   M9   &    M8   &    M9   &   M9--L0   &  M9 $\pm$ 1   \\
ROXs~42B b    &   L1--L2   &    M9   &    L0--L2   &   L1--L2   &  L1 $\pm$ 1   
\enddata
\end{deluxetable}

%\clearpage

\begin{deluxetable}{cccccc}
\tabletypesize{\small}
\tablewidth{0pt}
\tablecolumns{6}
\tablecaption{FW~Tau~b Emission Lines\label{tab:emission}}
\tablehead{
       \colhead{Species}            &    \colhead{Transition}                                &   \colhead{$\lambda$ (\AA)\tablenotemark{a}}        &  \colhead{Date}      &  \colhead{Line Flux\tablenotemark{b}}                 &    \colhead{EW\tablenotemark{c}}    \\
        \colhead{}      &           \colhead{}     &   \colhead{}            &  \colhead{(UT)}    &  \colhead{(10$^{-16}$ erg s$^{-1}$ cm$^{-2}$)}    &  \colhead{(\AA)}  
        }   
\startdata

$[$S II$]$     &   $^2$P$^o_{3/2}$--$^4$S$^o_{3/2}$    &   4068.6   &    2011-10-08    &  57 $\pm$ 12    &   $>$--101 $\pm$ 17  \\
                       &                                                                       &                  &    2013-01-02   &    8.1 $\pm$ 1.3 &   $>$--54 $\pm$ 8    \\ 
$[$O I$]$      &   $^1$S$_{0}$--$^1$D$_{2}$                  &   5577.3   &     2013-01-02   &  3.9 $\pm$ 1.0   &  $>$--18 $\pm$ 6   \\ 
$[$O I$]$       &  $^1$D$_{2}$--$^3$P$_{2}$                   &   6300.3   &      2011-10-08    &   11 $\pm$ 2  &  $>$ --11 $\pm$ 4    \\ 
                       &                                                                        &                  &       2013-01-02   &   10.4 $\pm$ 1.3  &  $>$--75 $\pm$ 8    \\ 
$[$OI$]$       &   $^1$D$_{2}$--$^3$P$_{1}$                   &   6363.8   &       2013-01-02   &  4.6 $\pm$ 1.0  &   $>$--24 $\pm$ 5   \\ 
H$\alpha$    &     $n$=3--2                                                   &  6562.7    &       2011-10-08   &   40 $\pm$ 4  &  $>$-67 $\pm$ 5     \\ 
                      &                                                                         &                  &       2013-01-02   &   35 $\pm$ 2  &  $>$--292 $\pm$ 12     \\ 
$[$S II$]$      &  $^2$D$^o_{5/2}$--$^4$S$^o_{3/2}$    &   6716.4   &       2011-10-08   &  6 $\pm$ 2   &  $\cdots$    \\ 
                      &                                                                         &                  &       2013-01-02   &  4.4 $\pm$ 0.6   &  $>$--24 $\pm$ 4    \\ 
$[$S II$]$       & $^2$D$^o_{3/2}$--$^4$S$^o_{3/2}$   &   6730.8   &       2011-10-08   & 9 $\pm$ 2   &  $>$--8 $\pm$ 4     \\ 
                       &                                                                        &                  &       2013-01-02   & 8.3 $\pm$ 0.6   &  $>$--57 $\pm$ 6     \\ 
?                    &   $\cdots$                                                     &   8286.4 $\pm$ 0.5       &   2013-01-02   & 3.8 $\pm$ 0.7  &  $>$--19 $\pm$ 4    \\ 
Ca II              &  $^2$P$^o_{3/2}$--$^2$D$_{3/2}$        &   8498.0   &       2013-01-02   & 6.5 $\pm$ 0.9     &  $>$--40 $\pm$ 5   \\ 
Ca II             &    $^2$P$^o_{3/2}$--$^2$D$_{5/2}$        &   8542.1   &       2013-01-02   &  3.5 $\pm$ 0.6  &  $>$--18 $\pm$ 4    \\ 
$[$Fe II$]$    &  $a$$^4$P$_{5/2}$--$a$$^4$F$_{9/2}$  &   8617.0   &       2013-01-02   &   2.3 $\pm$ 0.5   &  $>$--11 $\pm$ 4   \\ 
Ca II             &    $^2$P$^o_{1/2}$--$^2$D$_{3/2}$           &   8662.1   &       2013-01-02   &    5.8 $\pm$ 0.9   & $>$--35 $\pm$ 6   \\ 
$[$\ion{Fe}{2}$]$  & $a$$^4$D$_{7/2}$--$a$$^6$D$_{9/2}$  &   12567   &     2013-02-01  &  1.7 $\pm$ 0.4  &   --4.7 $\pm$ 1.3     \\ 
Pa$\beta$     &         $n$=5--3                                               &   12818.   &     2013-02-01   & 3.1 $\pm$ 0.4  &    --8.4 $\pm$ 1.0     \\ 
H$_2$           &   1--0 S(2)    &   20338.   &   2011-10-04  & 0.7 $\pm$ 0.2   &    --2.0 $\pm$ 0.6   \\ 
                       &                      &                   &   2013-02-01   & 1.9 $\pm$ 0.2  &    --5.9 $\pm$ 0.5   \\ 
H$_2$           &   1--0 S(1)    &   21218.   &   2011-10-04  &  4.6 $\pm$ 0.2  &    --13.6 $\pm$ 0.5   \\ 
                        &                      &                 &   2013-02-01   &  4.7 $\pm$ 0.2  &    --14.1 $\pm$ 0.6   \\ 
H$_2$           &   1--0 S(0)    &   22233.   &   2011-10-04  & 1.0  $\pm$ 0.2   &    --3.1 $\pm$ 0.5   \\ 
                        &                    &                    &   2013-02-01   & 0.61 $\pm$ 0.14  &    --1.8 $\pm$ 0.4   
\enddata
\tablenotetext{a}{Air wavelengths from the National Institute of Standards and Technology atomic line database (\citealt{Kramida:2012}) are used below 2 $\mu$m; vacuum wavelengths for $H_2$ from \citet{Tokunaga:2000p21771}
 are used above 2 $\mu$m.}
\tablenotetext{b}{The quoted uncertainties do not take into account the absolute flux calibration error, which we estimate to be 
$\sim$10\% for our SNIFS data.}
\tablenotetext{c}{Negative values indicate emission.  We do not detect continuum emission in our FW~Tau b optical spectra, so our measurements of line fluxes and equivalent widths represent lower limits.}

\end{deluxetable}

%\clearpage

\begin{deluxetable}{llcccc}
\tabletypesize{\small}
\tablewidth{0pt}
\tablecolumns{6}
\tablecaption{Fits to FW Tau b H$_2$ Emission Line Images\label{tab:fwspatfits}}
\tablehead{
        \colhead{Line}               &   \colhead{Region}   &  \colhead{FWHM$_\mathrm{maj}$}     &    \colhead{$a_\mathrm{max}$/$a_\mathrm{min}$}   &    \colhead{PA$_\mathrm{inst}$}  &    \colhead{PA$_\mathrm{sky}$}  
         }   
\startdata
\multicolumn{6}{c}{2011 Oct 04 UT} \\
\tableline
H$_2$ 1--0 S(2)      &  Emission Line                &    0.112     &    1.11  &      6    &  301   \\
                                   &  Continuum                      &     0.105     &    1.25   &    23    &  318   \\
                                   &  Cont-Subtracted            &     0.213     &    1.66    &  110  &     46   \\
H$_2$ 1--0 S(1)      &  Emission Line                &    0.134      &   1.09    &  138    &   73   \\
                                   &  Continuum                      &     0.108     &    1.24   &    25   &   320   \\
                                   &  Cont-Subtracted            &      0.195   &      1.38  &    124   &    60   \\
H$_2$ 1--0 S(0)      &  Emission Line                &     0.124      &   1.11    &   16   &   312   \\
                                   &  Continuum                      &      0.111     &    1.12   &   175 &     110   \\
                                   &  Cont-Subtracted            &      0.174    &     1.32  &     40 &     335   \\

\tableline
\multicolumn{6}{c}{2013 Feb 01 UT} \\
\tableline
H$_2$ 1--0 S(2)      &  Emission Line                &      0.156   &      1.15   &    44   &    249  \\
                                   &  Continuum                      &    0.149     &    1.20    &   72    &   277  \\
                                   &  Cont-Subtracted            &      0.241  &       1.64    &   13   &    218  \\
H$_2$ 1--0 S(1)      &  Emission Line                &   0.188      &   1.09     & 175    &  21  \\
                                   &  Continuum                      &    0.150     &    1.19    &   65    &   270  \\
                                   &  Cont-Subtracted            &     0.248   &      1.28   &   168   &   13  \\
H$_2$ 1--0 S(0)      &  Emission Line                &     0.156    &     1.11   &    60    &   265   \\
                                   &  Continuum                      &     0.161    &     1.26   &    71    &   276   \\
                                   &  Cont-Subtracted            &     0.198   &      1.50  &      3    &   209   
\enddata
\end{deluxetable}

\clearpage

\begin{figure}
  \vskip -.5in
  \begin{center}
  \resizebox{6.5in}{!}{\includegraphics{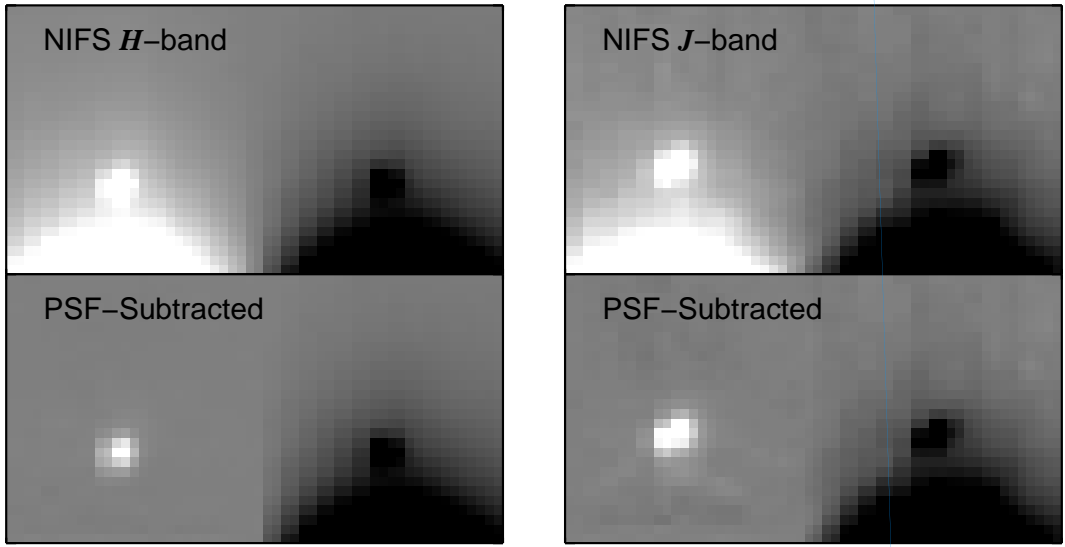}}
  \vskip -.5in
  \caption{Gemini-N/NIFS observations of ROXs~42B~b.  The upper images show examples of collapsed, pair-subtracted $H$- and $J$-band
  data cubes.  ROXs~42B~b is clearly visible but appears in the wing of the primary star's PSF (the separation is 1$\farcs$17).  
  The bottom panels show the same
  pair-subtracted images after fitting and subtracting a Moffat function to the contaminating flux for the positive image.     \label{fig:nifsimg} } 
\end{center}
\end{figure}

%\clearpage

\begin{figure}
  \vskip -.5in
  \begin{center}
  \resizebox{6.5in}{!}{\includegraphics{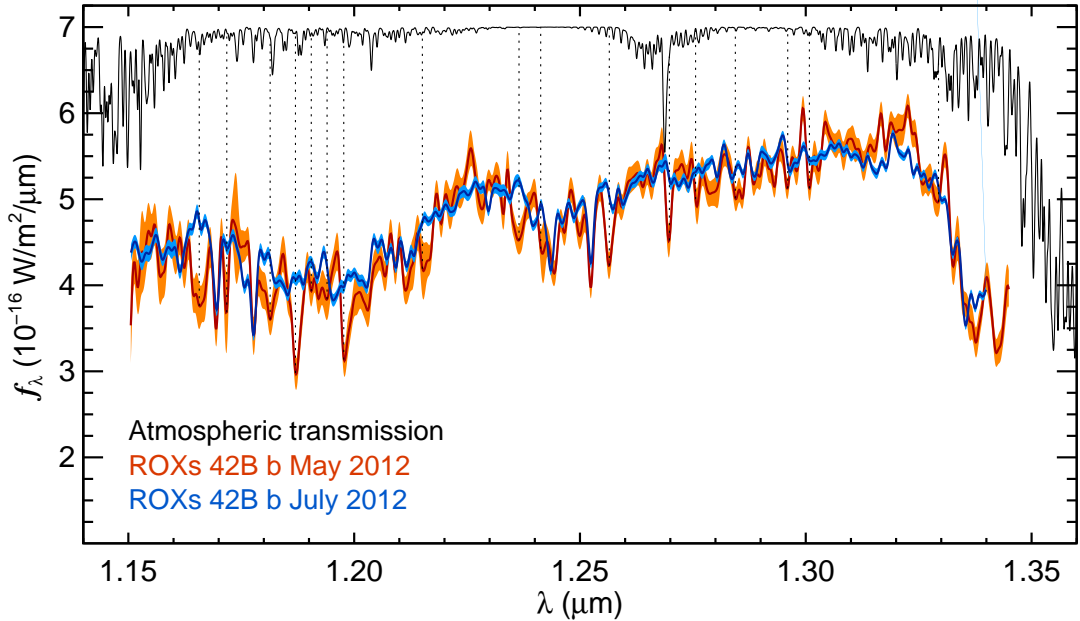}}
  \vskip -.5in
  \caption{Comparison of our telluric-corrected NIFS $J$-band spectra of ROXs~42B~b taken 
  in May 2012 (red) and July 2012 (blue).  The overall shape of the spectra agree but the
  detailed features differ.  Several of these discrepant lines (dotted black) in our lower-S/N 
  spectrum from May appear to be of telluric origin (solid black) and are probably caused by poor
  sky subtraction.  We therefore adopt the higher S/N July spectrum for this work.
  The spectra of ROXs~42B~b have
  been smoothed to a resolving power of 1000 and flux calibrated to 
  the measured $J$-band magnitude (16.99~$\pm$~0.07~mag; \citealt{Kraus:2014hk}).  
  1-$\sigma$ standard errors are shown as shaded orange and light blue.
  The transmission spectrum was generated by ATRAN (\citealt{Lord:1992p20487}) 
  assuming an airmass of 1.5 and
  a water vapor column of 1.6~mm (made available by Gemini Observatory).   \label{fig:roxs42_jcomp} } 
\end{center}
\end{figure}

\clearpage

\begin{figure}
  \vskip -1.7in
  \hskip -1.5in
  %\begin{center}
  \resizebox{9.5in}{!}{\includegraphics{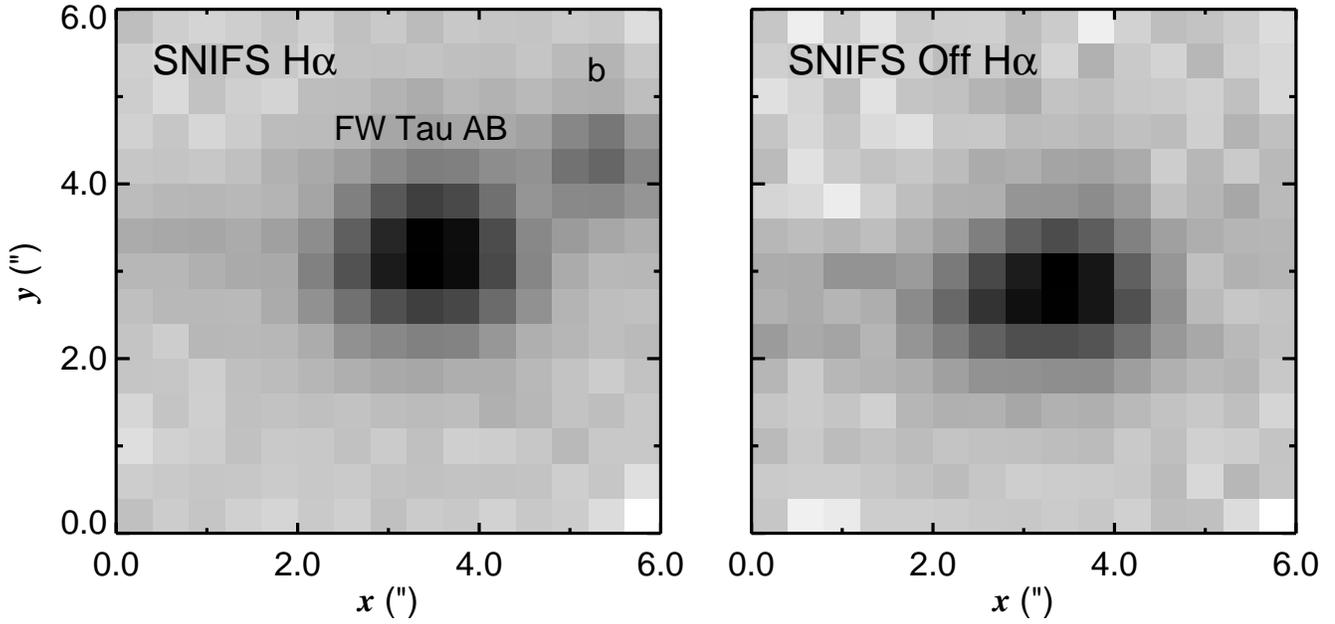}}
    \vskip -1.5in
  \caption{Seeing-limited SNIFS IFU images from 2013 of FW~Tau~AB+b on and off the H$\alpha$ line. 
  FW~Tau~b is detected in emission (left) but is not seen in the adjacent continuum (right).  The PSF is slightly elongated in
  the horizontal direction, probably from telescope windshake. \label{fig:snifsimg} } 
%\end{center}
\end{figure}

%\clearpage

\begin{figure}
  \vskip -.4in
  \begin{center}
  \resizebox{3.7in}{!}{\includegraphics{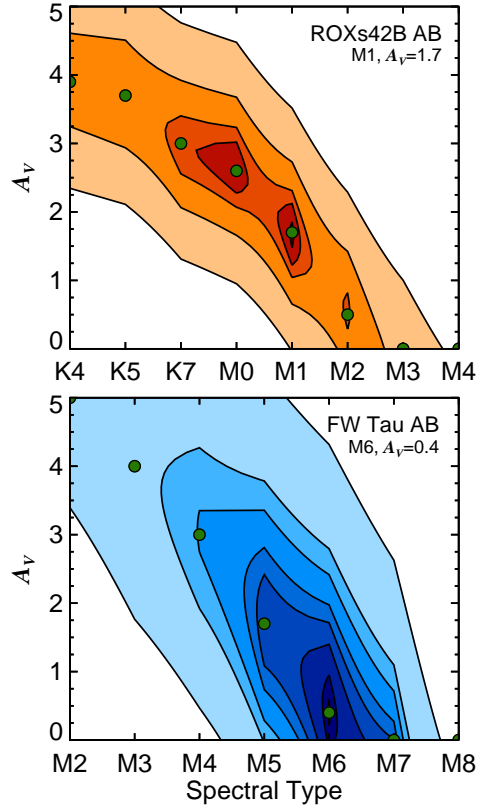}}
  \vskip -.4in
  \caption{$\chi^2$ contour plots showing the relative goodness-of-fit of spectral type and reddening for our
   SNIFS spectra of ROXs~42B~AB (top) and FW~Tau~AB (bottom).   Template spectra are from 
   \citet{Pickles:1998p17673} for K dwarfs and  \citet{Bochanski:2007p19461} for M dwarfs (see Section~\ref{sec:prispt}).
  Green circles indicate the best-fit reddening at each spectral type (displayed in Figure~\ref{fig:primaryspt}).  
  The best-matches to the 
   6000--8500~\AA \ region are \{M1~$\pm$~1, $A_V$=1.7$^{+0.9}_{-1.2}$~mag\} for ROXs~42B~AB
   and \{M6~$\pm$~1, $A_V$=0.4$^{+1.3}_{-0.4}$~mag\} for FW~Tau~AB.  
   Note the strong covariance between
   reddening and optical spectral type.
 \label{fig:chi2primary} } 
\end{center}
\end{figure}

\clearpage

\begin{figure}
  \vskip 1.in
  %\begin{center}
  \hskip .2in
  \resizebox{7.2in}{!}{\includegraphics{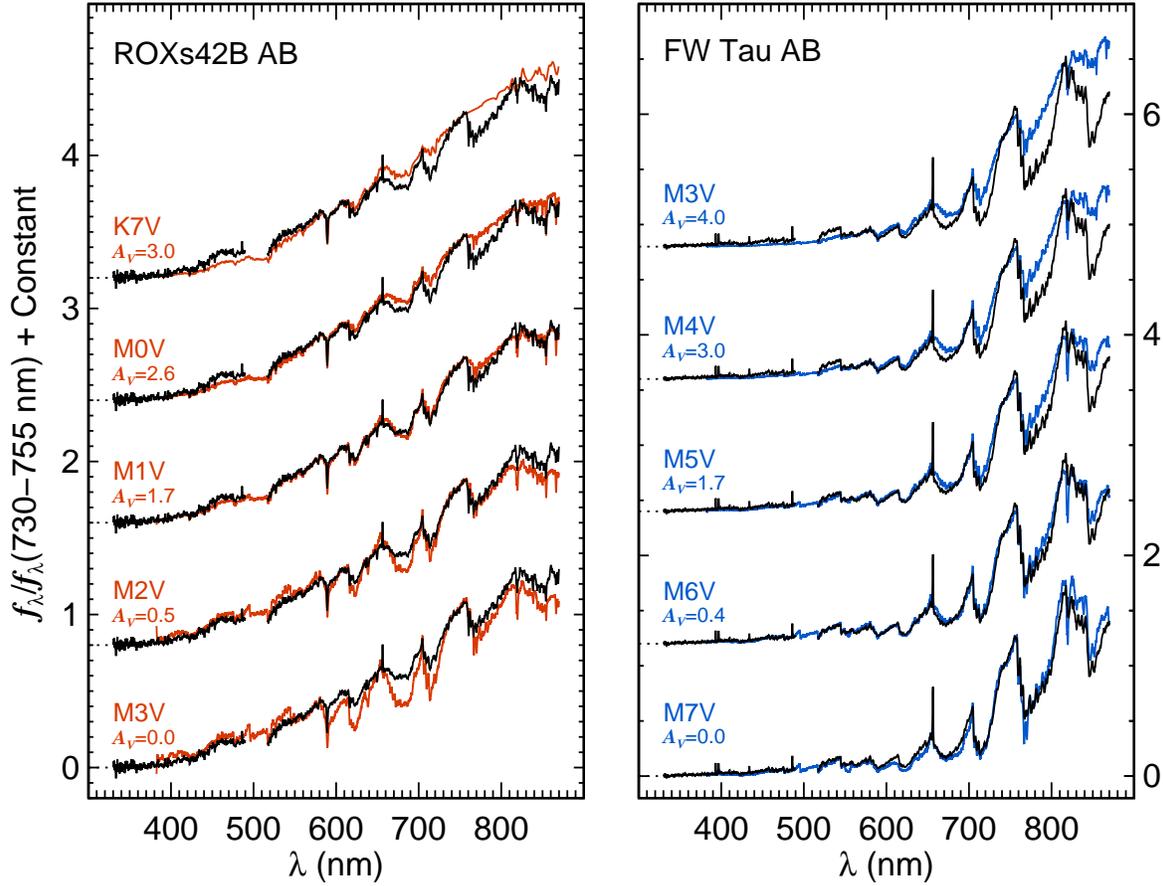}}
  \caption{Comparison of the best-fitting reddened optical templates to our SNIFS spectra ROXs~42B~AB and FW~Tau~AB 
  for each spectral type 
  (see Figure~\ref{fig:chi2primary}).  The best match is a moderately reddened M1 template for ROXs~42B~AB and 
  a slightly reddened M6 template for FW~Tau~AB.  The spectra are normalized between 7300--7550~\AA \ and
  offset by a constant.   \label{fig:primaryspt} } 
%\end{center}
\end{figure}

\clearpage

\begin{figure}
  \vskip 1.in
  \begin{center}
  \resizebox{4.2in}{!}{\includegraphics{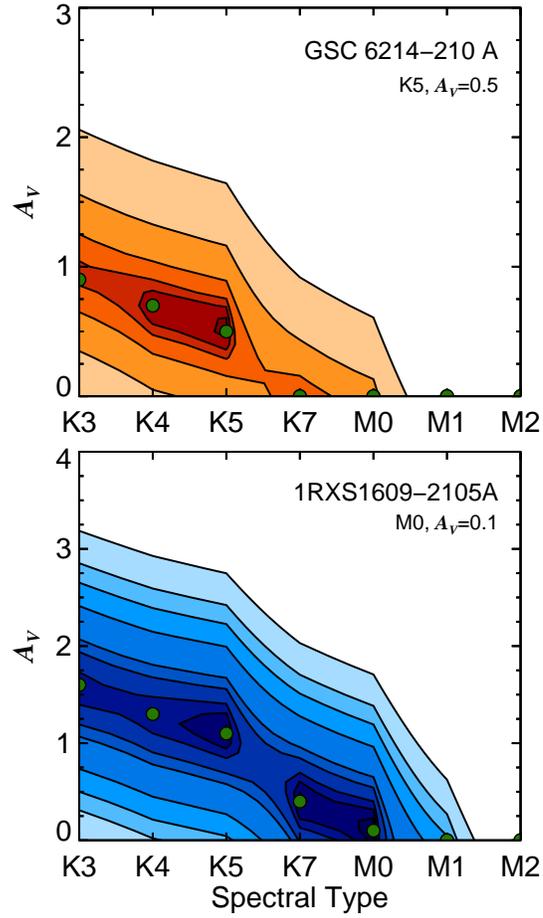}}
    \vskip -.5in
  \caption{$\chi^2$ contour plots for our reddening and spectral type fits to GSC~6214-210~A (top) and 1RXS~J1609--2105~A 
  (bottom).  See Figure~\ref{fig:primaryspt} for details.  The best fits are \{K5~$\pm$~1, $A_V$=0.5$^{+0.2}_{-0.5}$~mag\}
  for GSC~6214-210~A and \{M0~$\pm$~1, $A_V$=0.1$^{+0.3}_{-0.1}$~mag\}.     \label{fig:chi2_gsc_rxs_primary_spt} } 
\end{center}
\end{figure}

\clearpage

\begin{figure}
  \vskip 1.in
  %\begin{center}
    \hskip .2in
  \resizebox{7.2in}{!}{\includegraphics{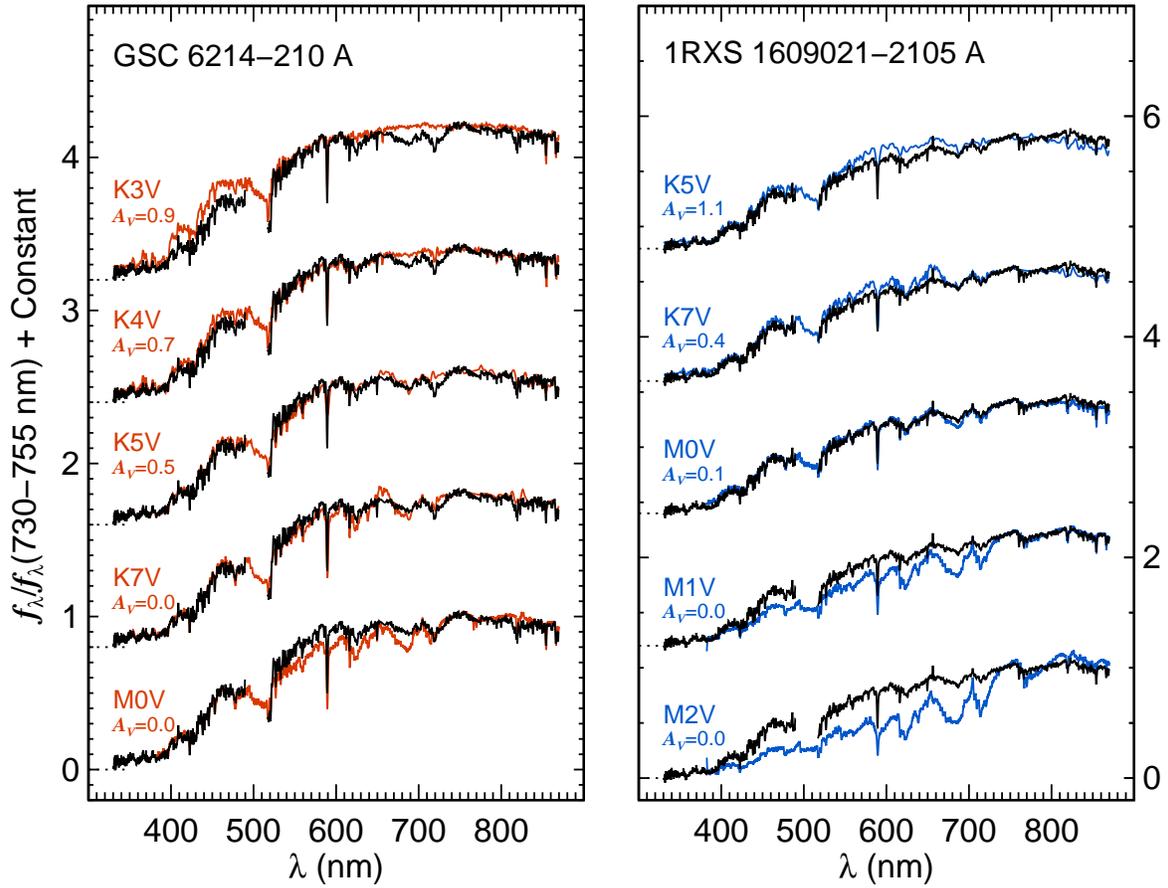}}
  \caption{SNIFS spectra of GSC~6214-210~A and 1RXS~J1609--2105~A compared to the best-fit reddened
  optical templates (see Figure~\ref{fig:chi2_gsc_rxs_primary_spt}).  
  A slightly reddened K5 spectrum is somewhat preferred over the unreddened
  K5 type we found for  GSC~6214-210~A in \citet{Bowler:2011p23014}.  The best match to 1RXS~J1609--2105~A
  is the M0 template. The spectra are normalized between 7300--7550~\AA \ and
  offset by a constant. \label{fig:primary_gsc_rxs_spt} } 
%\end{center}
\end{figure}

\clearpage

\begin{figure}
  \vskip -.5in
  \begin{center}
  \resizebox{6.5in}{!}{\includegraphics{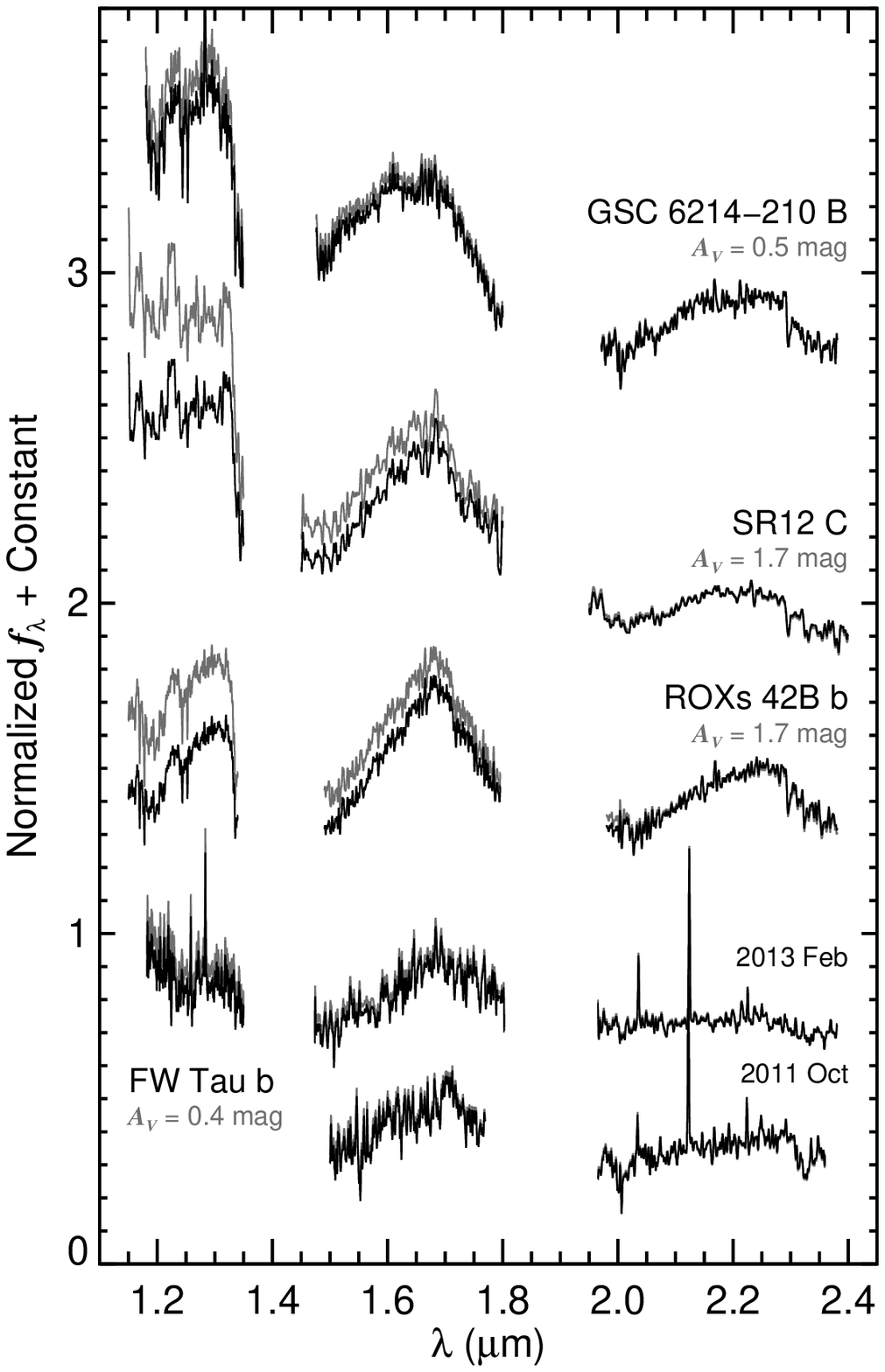}}
    \vskip -.8in
  \caption{Our near-infrared spectra of very young (2--10~Myr) low-mass (5--15~\Mjup) 
  companions.   The $J$- and $H$-bands
  of GSC~6214-210~B are from \citet{Bowler:2011p23014}, but the rest are new.  The spectra 
  are flux-calibrated to photometry from \citet{Ireland:2011p21592}, \citet{Kuzuhara:2011p21922}, and \citet{Kraus:2014hk},
  and are normalized to the 1.65--1.67~$\mu$m region.  Overplotted in gray are de-reddened spectra based on
  extinction values to the primary stars (normalized in $K$ band).  GSC~6214-210~B does not show Br$\gamma$
  emission, despite its very strong Pa$\beta$ emission.  The angular $H$-bands and weak $J$-band alkali absorption
  features in these objects are hallmarks of low gravity.  Our two epochs of NIR spectra for FW~Tau~b show a number of strong
  emission lines from H$_2$, Pa$\beta$, and $[$\ion{Fe}{2}$]$ (see Figure~\ref{fig:fwtauemission}), 
  but this object is otherwise relatively flat and featureless.  This may be a result of infrared veiling from a warm inner 
  disk or possibly an edge-on disk (see Section~\ref{sec:fwtaub}).    \label{fig:pmc_specfig} } 
\end{center}
\end{figure}

\clearpage

\begin{figure}
  \vskip -1in
  \hskip -.6in
  %\begin{center}
  \resizebox{8.5in}{!}{\includegraphics{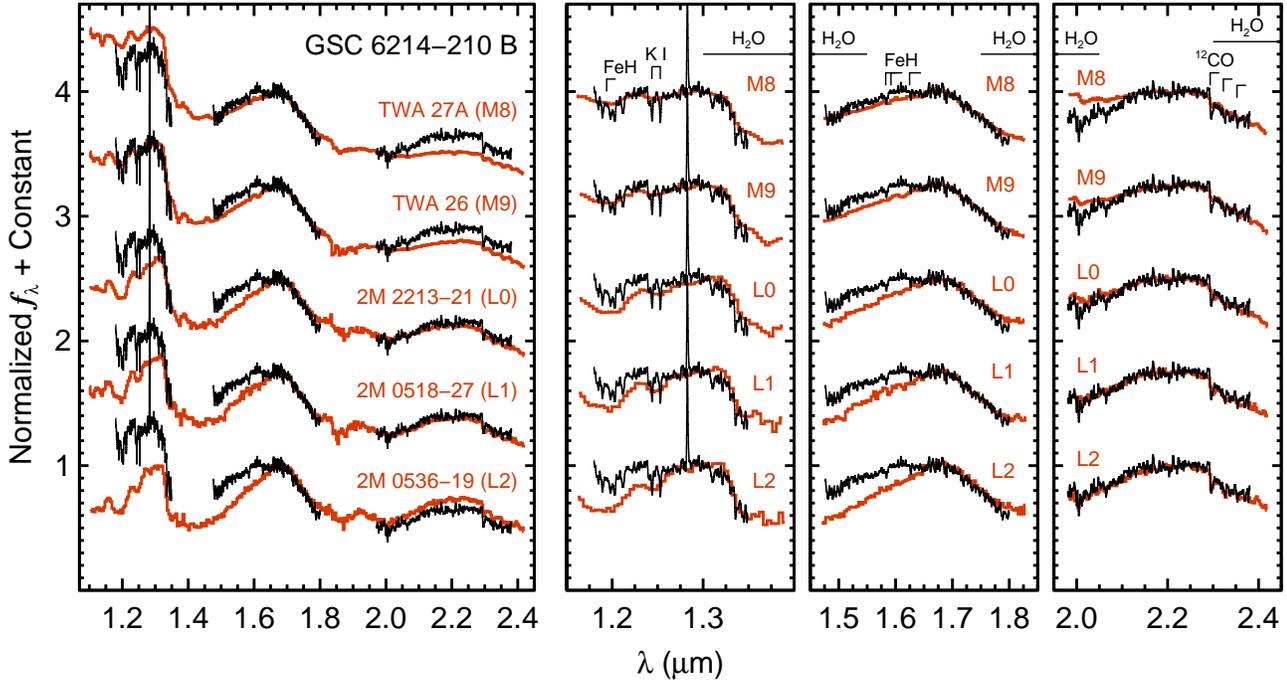}}
  \vskip -1.2in
  \caption{NIR spectral classification of GSC~6214-210~B (black) based on the very low-gravity sequence (red) defined by Allers \& Liu (2013, ApJ, submitted).  The left panel shows the flux-calibrated spectrum of GSC~6214-210~B de-reddened by $A_V$=0.5~mag and
  smoothed to $R$=1000.  The right three panels show the same comparison for the individual $J$, $H$, and $K$ bands.
  The SED (which is dominated by the accuracy of the photometry) is somewhat redder than the M9 template.  The $J$, $H$, and $K$ bands match the M9, M8, and L1--L2 templates, respectively; altogether we adopt  a NIR spectral type of M9.5~$\pm$~1.0.
  The spectra are normalized to the 1.66--1.68~$\mu$m regions for the first and third panels, and the 1.29--1.31~$\mu$m and 2.20--2.25~$\mu$m regions for the second and fourth panels, respectively.   
  Major absorption features from \ion{K}{1}, FeH, H$_2$O, and CO are labeled.
   \label{fig:gsc6214_vlgcomp} } 
%\end{center}
\end{figure}

%\clearpage

\begin{figure}
  \vskip -1in
  \hskip -.6in
  %\begin{center}
  \resizebox{8.5in}{!}{\includegraphics{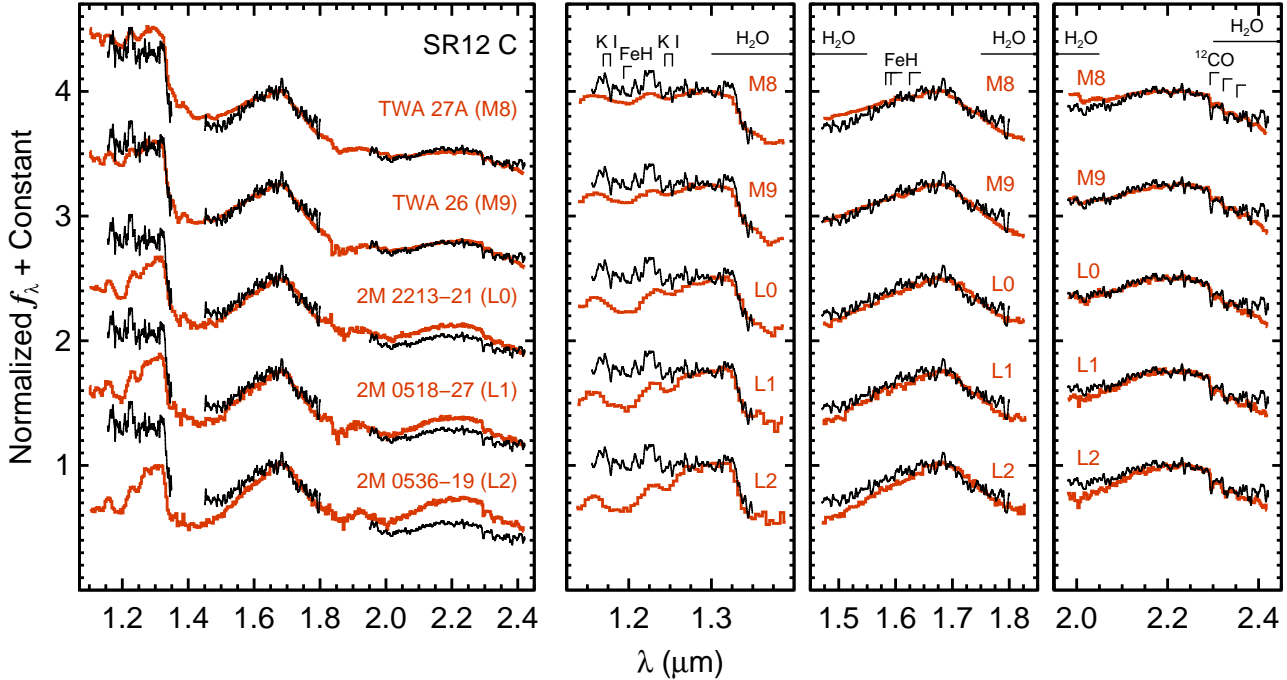}}
  \vskip -1.2in
  \caption{NIR spectral classification of SR~12~C (similar to Figure~\ref{fig:gsc6214_vlgcomp}).   The M9 template is an excellent fit
  to our spectrum.  The M8, M9, and M9--L0 templates best match the individual bandpasses (from left to right).  We adopt a NIR spectral type of M9.0~$\pm$~0.5, which is identical to the classification by \citet{Kuzuhara:2011p21922} based on a low-resolution spectrum.  Our IRTF moderate-resolution spectrum has been smoothed to $R$=500 and de-reddened by $A_V$=1.7~mag.   \label{fig:sr12c_vlgcomp} } 
%\end{center}
\end{figure}

\clearpage

\begin{figure}
  \vskip -1in
  \hskip -.6in
  %\begin{center}
  \resizebox{8.5in}{!}{\includegraphics{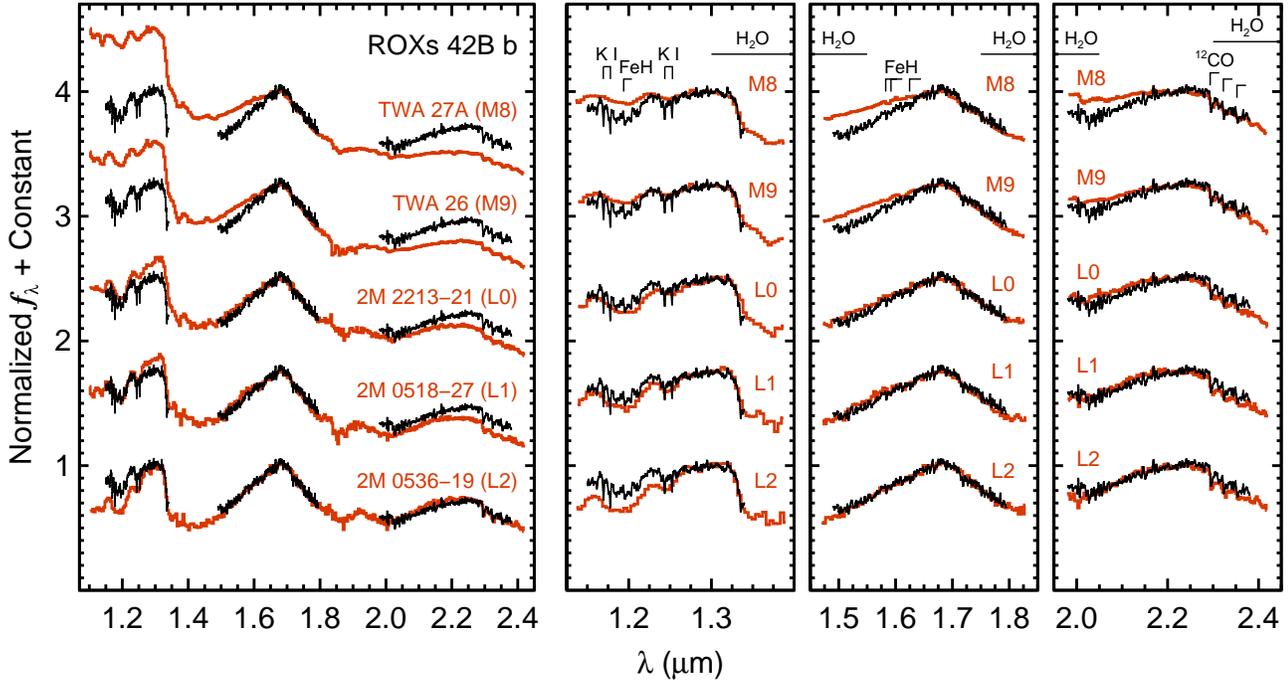}}
  \vskip -1.2in
  \caption{NIR spectral classification of ROXs~42B~b (similar to Figure~\ref{fig:gsc6214_vlgcomp}).  The SED of ROXs42B~b is slightly redder than the L1 template (left panel), and the best-matches to the individual bands (right panels) range from M9--L2.  
  Altogether, we adopt a NIR spectral type of L1.0~$\pm$~1.0.  Our flux-calibrated spectrum has been smoothed to $R$=1000 and de-reddened by $A_V$=1.7~mag.     \label{fig:roxs42_vlgcomp} } 
%\end{center}
\end{figure}

%\clearpage

\begin{figure}
  \vskip -1.in
  \begin{center}
  \resizebox{4.in}{!}{\includegraphics{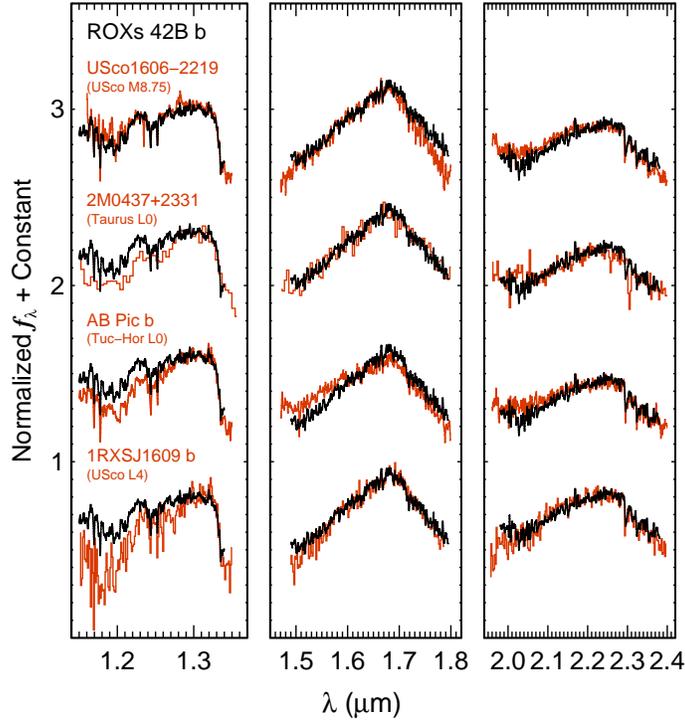}}
  \vskip -.8in
  \caption{Comparison of our de-reddened spectrum of ROXs~42B~b to other young substellar objects.
   Overall, ROXs~42B~b is most similar to the L0 Taurus member 2M0437+2331 and the $\sim$L4 Upper
   Scorpius companion 1RXS1609--2105 b in $H$ and $K$ bands, and the 
   M8.75 Upper Scorpius brown dwarf USco~1606--2219 in $J$-band.  Note that the $J$-band comparisons
   are the most prone to uncertainties in our reddening estimate of ROXS~42B~AB+b  
   ($A_V$=1.7$^{+0.9}_{-1.2}$~mag).  The comparison spectra of USco~1606--2219, 
   and AB~Pic~b are from \citet{Lodieu:2008p8698} and \citet{Bonnefoy:2010p20602}, respectively.
   The 1RXS1609--2105 b spectrum is from \citet{Lafreniere:2008p14057} and \citet{Lafreniere:2010p20986}.
   We assume zero reddening to these comparison objects.
    \label{fig:roxs42bb_comp} } 
\end{center}
\end{figure}

\clearpage

\begin{figure}
  \vskip -1.in
  \begin{center}
  \resizebox{6.5in}{!}{\includegraphics{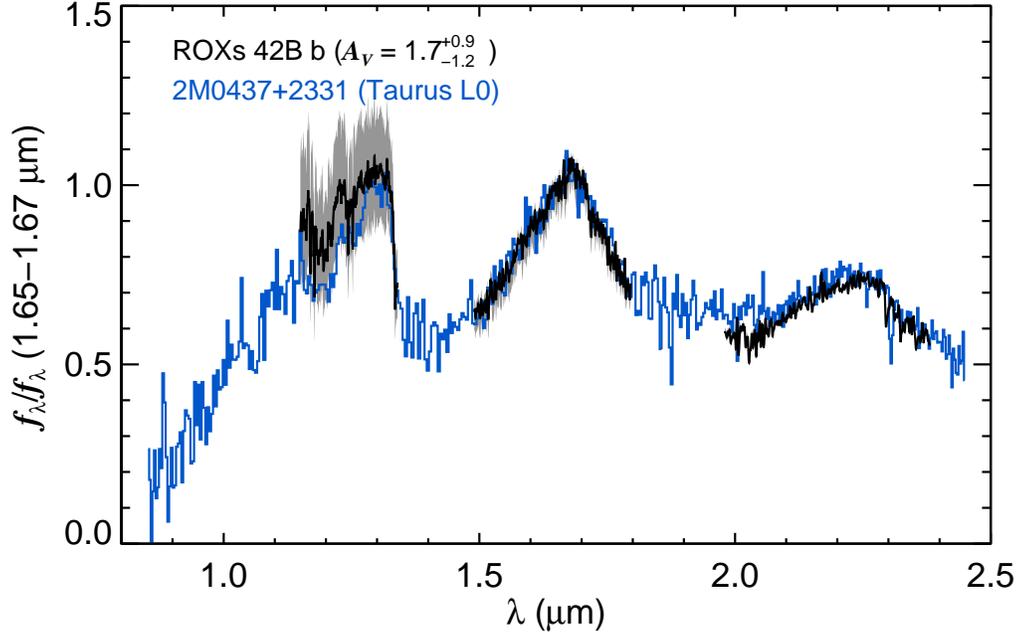}}
  \vskip -.5in
  \caption{Our de-reddened spectrum of ROXs~42B~b (black) bears a close resemblance to 
   the 4--7~\Mjup \ L0 Taurus member 2M0437+2331 (blue).    The gray shaded region shows
   the range of extinction values for ROXs~42B~b based on our fits to the primary (Section~\ref{sec:prispt}).
   Both spectra are normalized to the 1.65--1.67~$\mu$m region.
     \label{fig:roxs42bb_2m0437_comp} } 
\end{center}
\end{figure}

%\clearpage

\begin{figure}
  \vskip -0.in
  \hskip .5in
  %\begin{center}
  \resizebox{7.5in}{!}{\includegraphics{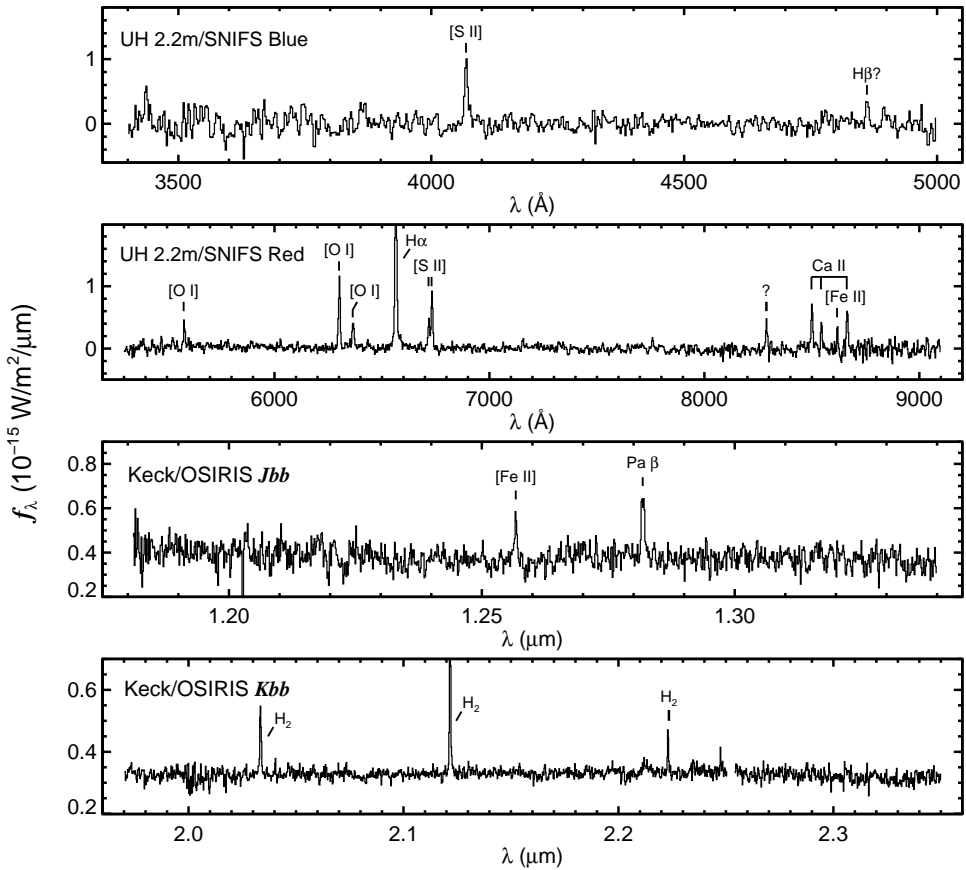}}
  \vskip -.6in
  \caption{Optical through near-infrared emission line spectrum of FW~Tau~b.  Our SNIFS spectrum
  (2013 observations; top two panels) shows a wealth of permitted and forbidden emission
  lines, but does not reach the continuum level.  Our $J$- and $K$-band spectra (2013 observations; bottom two panels) also show
  emission lines associated with accretion and shock excitation, as well as mostly flat continuum.  The spectra are absolutely flux-calibrated and prominent emission lines are labeled.  Line fluxes and equivalent widths are listed in Table~\ref{tab:emission}.  \label{fig:fwtauemission} } 
%\end{center}
\end{figure}

\clearpage

\begin{figure}
  %\vskip -2.in
  \begin{center}
  \hskip -0.2in
  \resizebox{5.in}{!}{\includegraphics{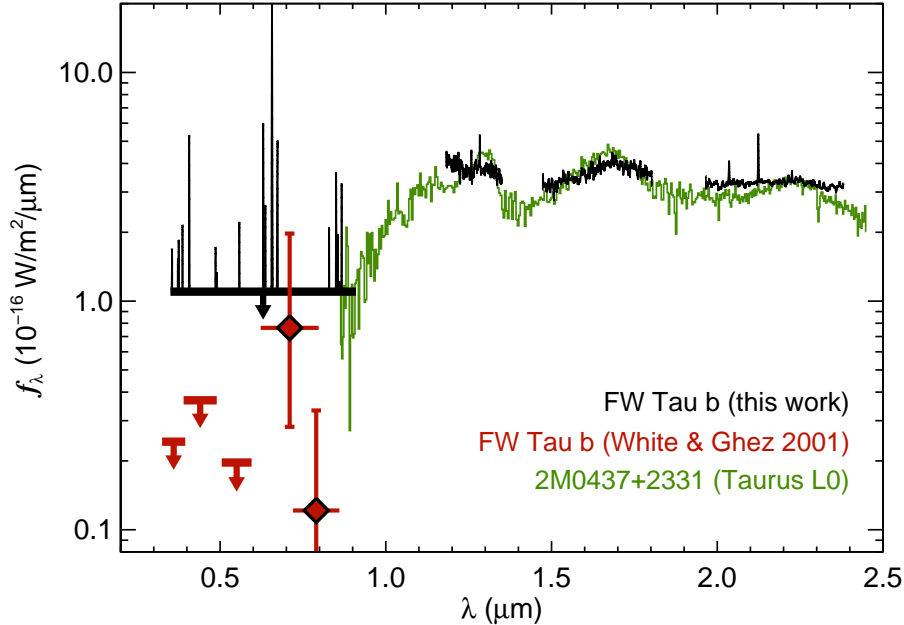}}
  \caption{Optical through NIR observations of FW~Tau~b.  Our flux-calibrated 2013 optical spectrum is shown
  with the 1-$\sigma$ continuum upper limit  as a thick black horizontal bar at 
  1.1$\times$10$^{-16}$ W m$^{-2}$ $\mu$m$^{-1}$ (the SNIFS data have been smoothed here).
  Overplotted in red are the \citet{White:2001p21137}
  $HST$ detections ($F675W$, $F814W$) and upper limits ($F336W$, $F439W$, $F555W$) as converted
  to Johnson-Cousins filter system.  Note that the $R$-band detection includes strong H$\alpha$
  emission.  For comparison, we also plot our flux-calibrated spectrum of 2M0437+2331 (no relative scaling is
  applied to either spectrum).  Both 2M0437+2331  and FW~Tau~b are members of Taurus so have similar
  ages and distances.  Their apparent NIR magnitudes and colors are similar, but their spectra are somewhat
  different.  
   \label{fig:fwtaub_specsed} } 
\end{center}
\end{figure}

%\clearpage

\begin{figure}
  \vskip -.3in
  %\hskip 0.8in
  \begin{center}
  \resizebox{5.7in}{!}{\includegraphics{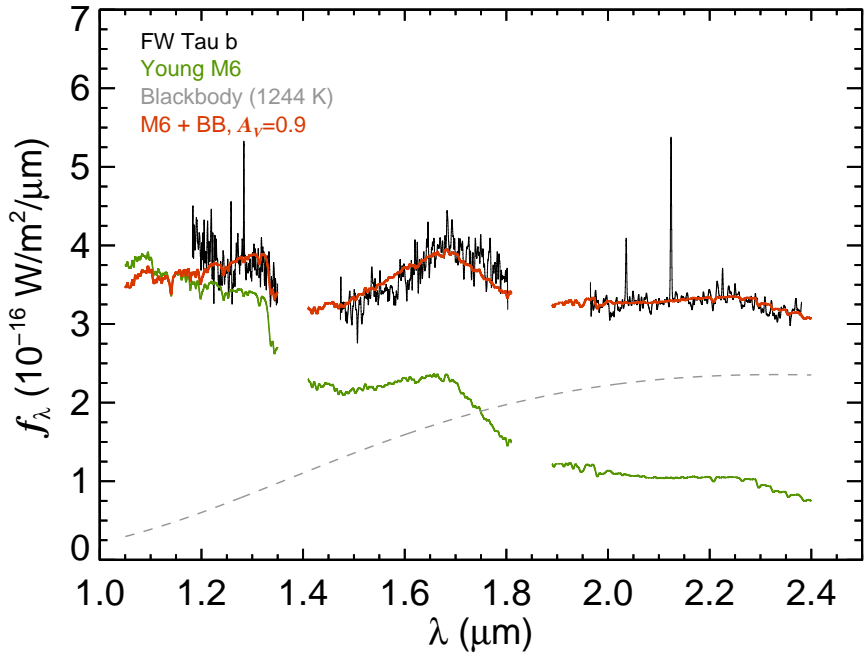}}
  \vskip -.1in
  \caption{Our NIR spectrum of FW~Tau~b is reasonably well-fit with a modestly reddened late-M dwarf 
  photosphere (green) and a hot blackbody component (dashed gray).  The total emission (red)
  reproduces the shape of our flux-calibrated spectrum from 1.2--2.4~$\mu$m.  
  Among the very low-gravity spectral templates from \citet{Allers:2013p25314}, the
  young M6 brown dwarf TWA~8B (originally from \citealt{Allers:2009p18424})
  provides the best match together with a 1244~K blackbody and modest reddening of $A_V$=0.9~mag.
  This suggests that the edge-on disk hypothesis is plausible, though more observations will be needed
  to distinguish these scenarios.
  Note that emission lines were removed from our spectrum of FW~Tau~b for this exercise.
   \label{fig:fwtaub_fitbb} } 
\end{center}
\end{figure}

\clearpage

\begin{figure}
  \vskip 1.in
  \hskip -1in
  %\begin{center}
  \resizebox{8.5in}{!}{\includegraphics{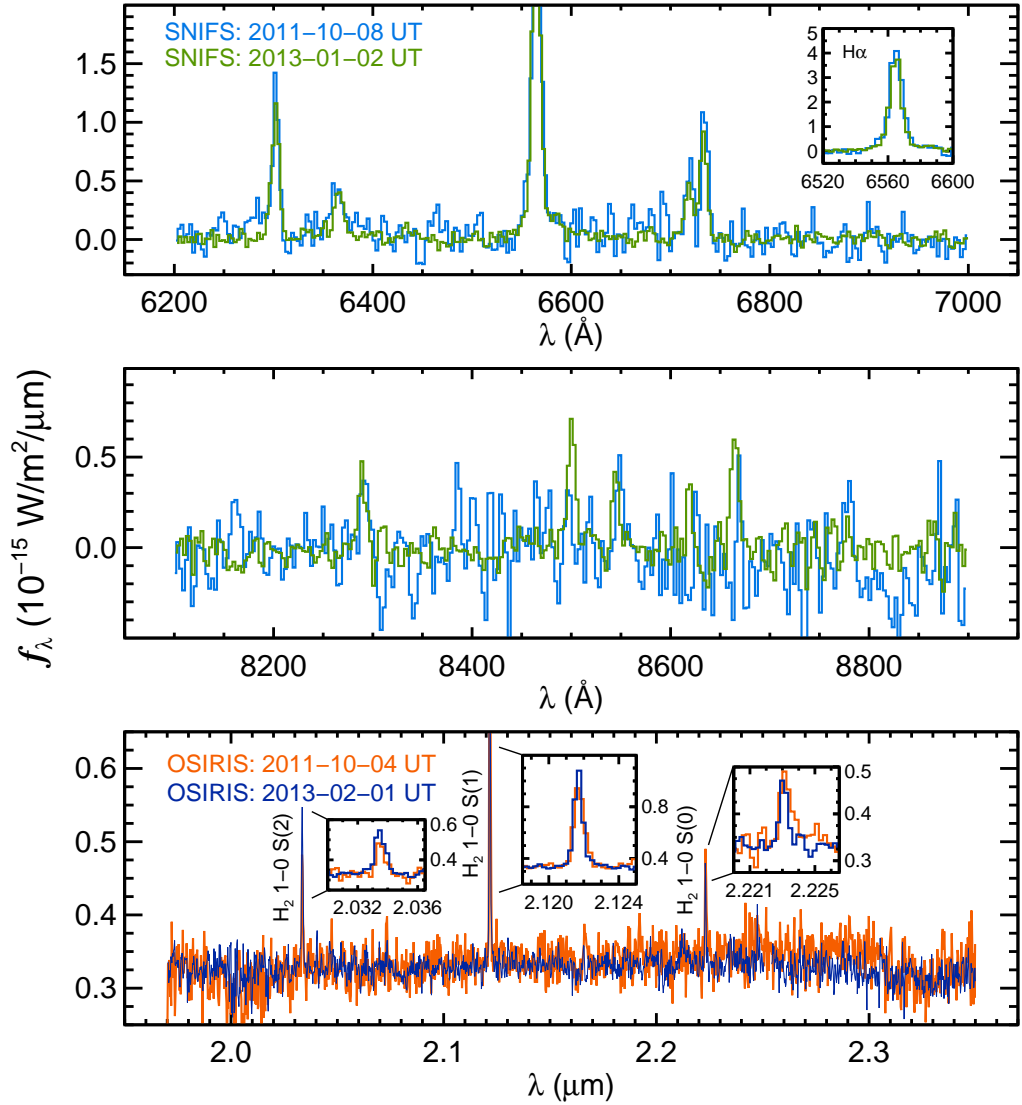}}
  \caption{Comparison of our two epochs of flux-calibrated optical and NIR spectroscopy of FW~Tau~b.  
  Most of the emission lines are constant over $\sim$1.5 years, with only two showing evidence
  of variability at the $>$3-$\sigma$ level: H$_2$~1--0~S(2) and $[$\ion{S}{2}$]$~$\lambda$4067.
  Insets show the relative intensities of the strongest lines in our data.
  \label{fig:snifs_var_fwtaub} } 
  %\end{center}
\end{figure}

\clearpage

\begin{figure}
  \vskip .5in
  \hskip 1.2in
  \resizebox{6.7in}{!}{\includegraphics{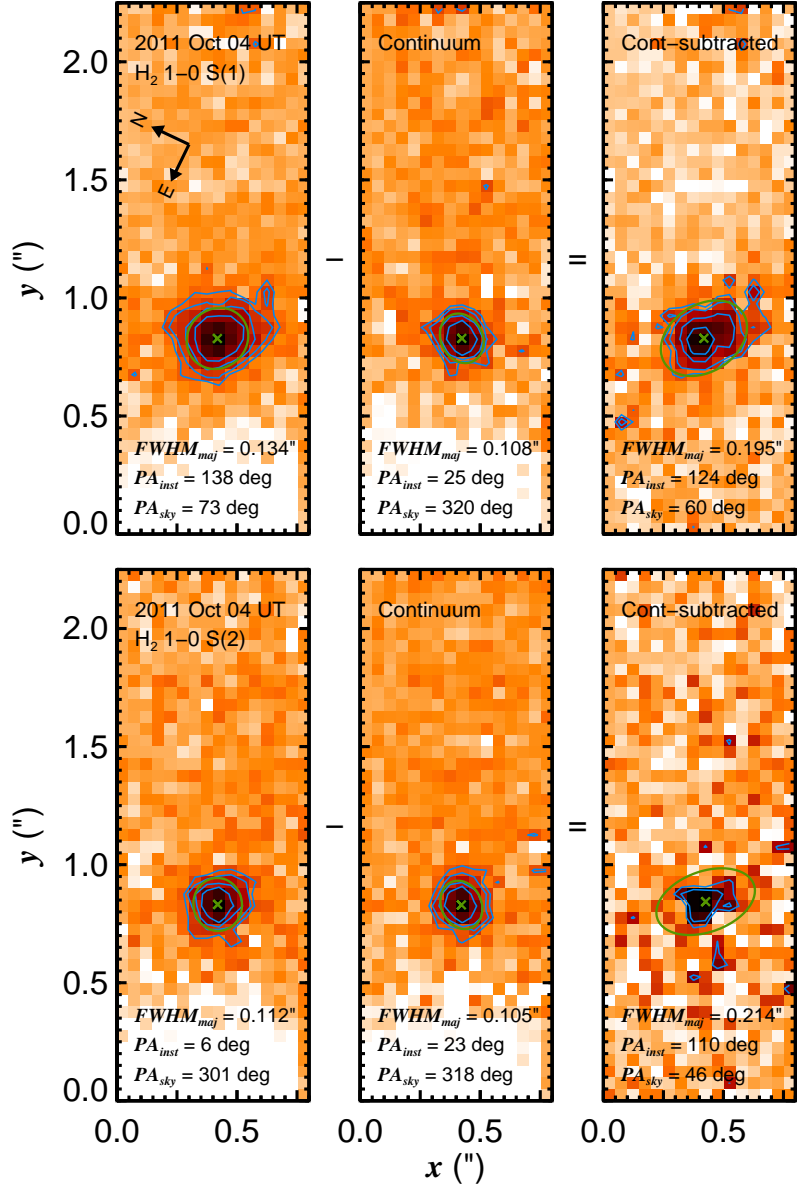}}
  \vskip -2.5in
  \caption{2011 Keck/OSIRIS H$_2$ images of FW~Tau~b.  Each frame has been registered and coadded
   from our nodded observations.
  The top panels show 2.122~$\mu$m H$_2$~1--0~S(1) images of the emission line (left), continuum emission 
  immediately adjacent to the line (center), and continuum-subtracted residuals (right).  Blue contours represent
  \{3, 5, 10, 20\}-$\sigma$ levels relative to the background rms.  Green ellipses and crosses show the FWHM and
  centers from fitting a
  2D elliptical Gaussian to the source.  The FWHM of the major axis, instrumental position angle (PA), and sky PA
  are listed, and a compass rose is shown for orientation.  
  The FWHM of the continuum-subtracted H$_2$~1--0~S(1) line is about twice as large as the continuum image,
  indicating FW~Tau~b is marginally resolved along that axis.  The extended emission appears asymmetric 
  in the +240$^{\circ}$ direction.  The bottom panels are identical for the 2.034~$\mu$m H$_2$~1--0~S(2)
  emission line.  It shows a similar slight, but significant, extension in the +226$^{\circ}$ direction.
  The morphology is similar to other H$_2$ jets and probably originates from outflow material,
  although the structure is not kinematically resolved in our data.
  \label{fig:fwtaub_h2_2011} } 
\end{figure}

\clearpage

\begin{figure}
  \vskip .5in
  \hskip 1.2in
  \resizebox{6.7in}{!}{\includegraphics{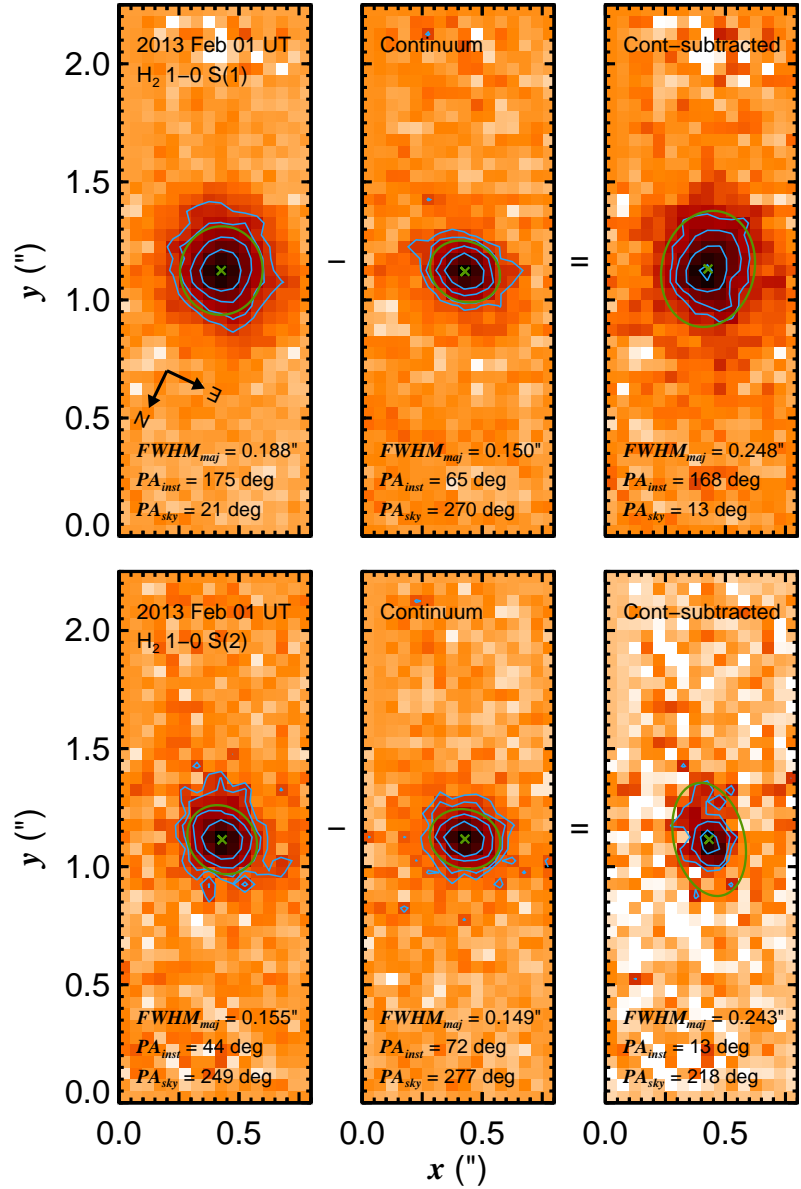}}
  \vskip -2.6in
  \caption{H$_2$ images from our 2013 OSIRIS integral field observations of FW~Tau~b.
  The emission lines, contour plots, and Gaussian fits are identical to Figure~\ref{fig:fwtaub_h2_2011}.
  Seeing conditions and AO correction were substantially worse compared to our 2011 data,
  which is evident by the continuum FWHM measurements.  Nevertheless, the continuum-subtracted
  2.034~$\mu$m H$_2$~1--0~S(2) line show similar extension in the +218$^{\circ}$ direction 
  and is broadly consistent with our 2011 observations.  Note that the large difference in North orientation
  is a result of the different detector orientations (perpendicular here instead of parallel to the binary PA).
  \label{fig:fwtaub_h2_2013} } 
%\end{center}
\end{figure}

\clearpage

\begin{figure}
  \vskip .5in
  \hskip 1.2in
  \resizebox{4.8in}{!}{\includegraphics{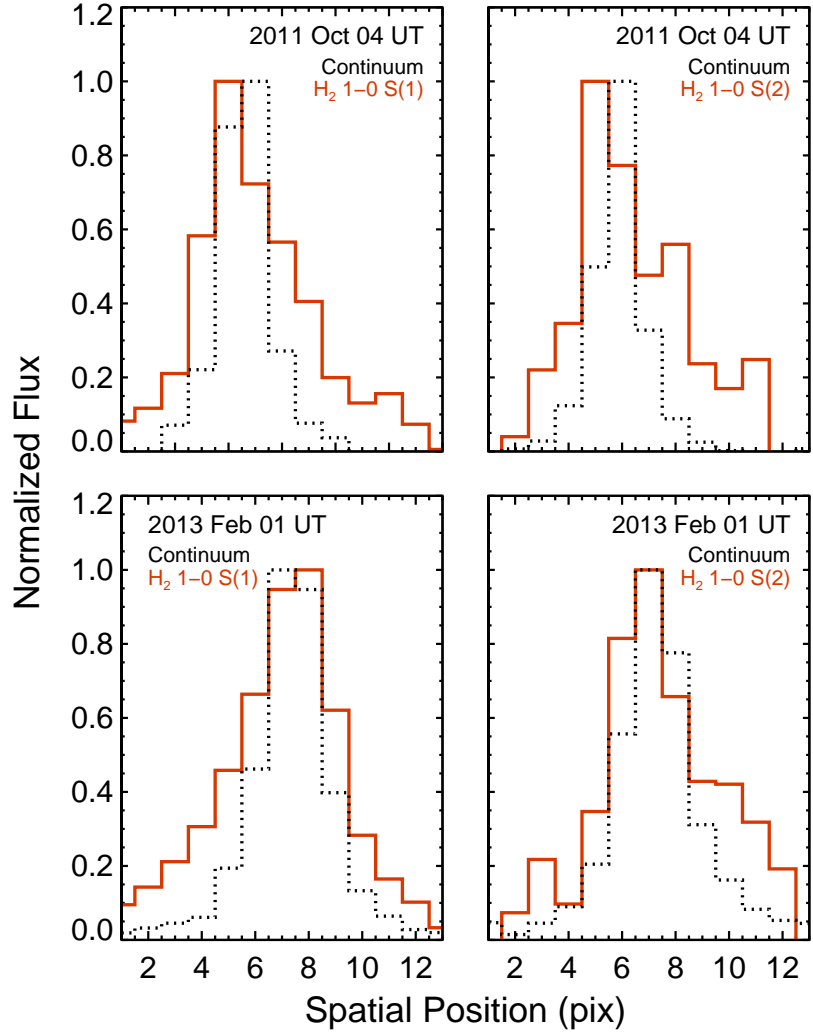}}
  \vskip -.1in
  \caption{ H$_2$~1--0~S(1) and 1--0~S(2) continuum-subtracted spatial line profiles (red) compared to adjacent continuum spatial profiles (dotted black).  The panels show the spatial cross section of each emission line along its major axis of the continuum-subtracted frames from 2011 (Figures~\ref{fig:fwtaub_h2_2011}) and 2013 (Figure~\ref{fig:fwtaub_h2_2013}).  In each case the normalized line profile appears extended compared to adjacent continuum, indicating resolved structure along that direction, most likely from outflow activity.  Each cross section is oriented so that the SW direction of the cut is to the right.  The OSIRIS plate scale is 50~mas per spaxel.
  \label{fig:fwtaub_h2_profiles} } 
%\end{center}
\end{figure}

\clearpage

\end{document}